\documentclass[aps,prc,twocolumn,showpacs,preprintnumbers,
                          nofootinbib,float,superscriptaddress,longbibliography]{revtex4-1}
\usepackage{graphicx, fancybox}
\usepackage{amsmath,amssymb}
\usepackage[colorlinks=true, pdfstartview=FitV, linkcolor=red,
		     citecolor=blue, urlcolor=blue]{hyperref}
\usepackage[usenames, dvipsnames]{color}
\usepackage{soul}
\usepackage{url}

\usepackage[normalem]{ulem}

\begin{document}

\title{Hydrodynamic Predictions for Pb+Pb Collisions at 5.02 TeV}

\author{Scott McDonald}
\affiliation{Department of Physics, McGill University, 3600 University
Street, Montreal,
QC, H3A 2T8, Canada}

\author{Chun Shen}
\affiliation{Department of Physics, McGill University, 3600 University
 Street, Montreal,
 QC, H3A 2T8, Canada}
 
\author{Fran\c{c}ois Fillion-Gourdeau}
\affiliation{Universit\'e du Qu\'ebec, INRS-\'Energie, Mat\'eriaux et
T\'el\'ecommunications, Varennes, Qu\'ebec, Canada J3X 1S2}
\affiliation{Institute for Quantum Computing, University of Waterloo, Waterloo, Ontario, Canada, N2L 3G1}

\author{Sangyong Jeon}
 \affiliation{Department of Physics, McGill University, 3600 University
 Street, Montreal,
 QC, H3A 2T8, Canada}
 
 \author{Charles Gale}
 \affiliation{Department of Physics, McGill University, 3600 University
 Street, Montreal,
 QC, H3A 2T8, Canada}

\begin{abstract}
Predictions and comparisons of hadronic flow observables for Pb+Pb collisions at 2.76 TeV and 5.02 TeV are presented using a hydrodynamics + hadronic cascade hybrid approach. Initial conditions are generated via a new formulation of the IP-Glasma model and then evolved using relativistic viscous hydrodynamics and finally fed into transport cascade in the hadronic phase. The results of this work show excellent agreement with the recent charged hadron anisotropic flow measurements from the ALICE collaboration \cite{Adam:2016izf} of Pb+Pb collisions at 5.02 TeV. Event-by-event distributions of charged hadron $v_n$, flow event-plane correlations, and flow factorization breaking ratios are compared with existing measurements at 2.76 TeV, and are predicted at 5.02 TeV. Further predictions of identified hadron observables (for both light and multi-strange hadrons), such as $p_T$-spectra and anisotropic flow coefficients, are presented. 
\end{abstract}

\maketitle
\date{\today}

\section{Introduction}

Relativistic heavy-ion collisions conducted at the Relativistic Heavy-Ion Collider (RHIC) and the Large Hadron Collider (LHC) create a deconfined state of quarks and gluons, known as quark-gluon plasma (QGP), at extreme temperatures and densities. Astonishingly, the strongly-coupled nature of the QGP exhibits nearly perfect fluid behaviour in experimental measurements. On the one hand, such collective phenomena are very hard to study from first principles due to the strong coupling that restricts our ability to use perturbative QCD techniques. On the other hand, the wealth of experimental data coupled with macroscopic phenomenological models can offer us reliable tools to quantitatively constrain the transport properties of the QCD matter. The measured momentum distributions of produced hadrons, namely particle transverse momentum spectra and the Fourier coefficients of their azimuthal distributions (known as anisotropic flow coefficients $\{v_n\}$), show stringent power to extract the transport properties of QGP \cite{Gale:2013da, Heinz:2013th}. In the spirit of precisely mapping out properties of QGP, hadronic observables in Pb+Pb collisions at $\sqrt{s} = 5.02$ TeV center-of-mass energy were recently measured at the LHC \cite{Adam:2016izf}. 

In this work, we will confront this new data with a fully integrated state-of-the-art theoretical framework. For initial conditions, we rely on a new formulation of the IP-Glasma model, which provides realistic event-by-event fluctuations and non-zero pre-equilibrium flow at the early stage of heavy-ion collisions. 
Individual collision systems are evolved using relativistic hydrodynamics with non-zero shear and bulk viscosities \cite{Ryu:2015vwa}. As the density of the system drops, fluid cells are converted into hadrons and further propagated microscopically using a hadronic cascade model \cite{Bass:1998ca,Bleicher:1999xi}.
A quantitative description of various hadronic flow observables in Pb+Pb collisions 2.76 TeV is achieved within this framework. Regarding this as a base point, we extend our approach to study the dynamical evolution of Pb+Pb collisions at the higher 5.02 TeV.
We will provide quantitative predictions for the flow observables of identified particles, namely particle spectra, mean $p_T$, and $p_T$-differential $v_{2,3,4}\{\mathrm{SP}\}(p_T)$, thus filling a void in the current literature of flow predictions in Pb+Pb collisions at 5.02 TeV \cite{Niemi:2015voa,Noronha-Hostler:2015uye,Ma:2016fve}.  The event-by-event distributions of charged hadron $v_n$ are studied, for the first time, within such a hybrid approach at the two LHC collision energies. The flow event-plane correlations and flow factorization breaking ratios with IP-Glasma initial conditions are also investigated and compared with the existing ATLAS and CMS measurements at 2.76 TeV. Predictions of these correlation observables at 5.02 TeV are presented. All of these flow observables at the two LHC energies will help us to verify the validity of the hydrodynamic model framework and, more importantly, to set stronger constraints on the extraction of transport properties of the QGP, especially their temperature dependence. 

The paper is organized as follows. Section II will provide an overview of the theoretical framework in three stages, beginning with a discussion on the new implementation of IP-Glasma and its underlying physics. Section IIB will describe the centrality selection procedure that is employed for this work, and section IIC will briefly introduce the details of the hydrodynamic and hadronic cascade simulations. The latter section will focus on the details of the implementations and the parameters used as the physics of these models has been discussed in great detail elsewhere \cite{Ryu:2015vwa,Ryu:2017qzn}. The procedure of our flow analysis is explained in Section IID.  In Section III, we start the phenomenology discussion from inclusive charged hadron observables to identified particle flow coefficients. Their centrality and $p_T$-dependence results are compared to existing experimental measurements at 2.76 TeV. Predictions are made at the higher 5.02 TeV. Event-by-event flow distribution and event-plane correlations are studied in Sec. IIIC at the two LHC collision energies. Finally, conclusions will be outlined in section IV. In the appendix, effects from out-of-equilibrium correlations and hadronic scatterings in particle flow coefficients and their correlation functions are discussed. 

\section{Theoretical Framework}

\subsection{The IP-Glasma Model}
Historically, the initial state of heavy ion collisions was dominated by geometric models such as the Monte Carlo Glauber model. In recent years, however, QCD based models that include saturation physics have come to define the standard in the field \cite{PhysRevC.94.024907}. IP-Glasma, in particular, has proven itself to be an excellent model of the initial state.\footnote{The EKRT model is another such saturation based model. An overview of EKRT, as well as a brief comparison with the IP-Glasma model, can be found in Ref. \cite{Niemi:2015qia}.} First implemented in Refs. \cite{Schenke:2012fw, Schenke:2012wb}, IP-Glasma is based on the IP-Sat model presented in \cite{Kowalski:2003hm}, and the glasma model \cite{Krasnitz:1998ns, Krasnitz:1999wc, Krasnitz:2001qu, Krasnitz:2000gz}. It includes saturation physics as well as sub-nucleonic color charge fluctuations that give the model its trademark ``spiky" initial conditions.  This paper presents a new numerical implementation of IP-Glasma, the results of which will be presented in the next section.
Numerical convergence of the energy density as a function of the lattice spacing  was confirmed for relevant switching times, and the results were consistent with those of Ref. \cite{Fukushima:2011nq}.  The simulations presented here utilize a lattice spacing of $dx=dy=0.1$ GeV$^{-1}$, which is sufficient for convergence.

The IP-Glasma model begins by sampling nucleons from a Woods-Saxon distribution,
\begin{equation}
\rho(r)/\rho_0=\frac{1}{1+\exp \left(\frac{r-R}{a} \right)}
\end{equation}
where $R$ is the nuclear radius, and $a$ is the nuclear skin depth.  For \textsuperscript{208}Pb, these parameters are set to $R = 6.62$\,fm and $a = 0.546$\,fm \cite{DeJager:1987qc}.\footnote{We checked that the initial eccentricities change by only a few percent if spatial configurations which includes nucleon-nucleon correlations \cite{Alvioli:2009ab} are used. Similar findings were shown in the MC-Glauber and the MCKLN models \cite{Shen:2015qta}.} Once the spatial distribution of nucleons is sampled and projected onto the transverse plane, the saturation scale is determined through the IP-Sat framework \cite{Kowalski:2003hm}. The criterion for the saturation scale comes from requiring that the exponent of the Glauber-Mueller dipole cross section given by
\begin{equation}
\frac{d\sigma_{qq}}{d^{2}b}= 2\left[1-\exp{\left(-\frac{\pi^{2}}{2N_{c}}r^{2}\alpha_{s}(\mu^{2})xg(x,\mu^{2})T(b)\right)} \right] 
\end{equation}
equals 1/2 as
\begin{equation}
\left(\frac{\pi^{2}}{2N_{c}}r^{2}\alpha_{s}(\mu^{2})xg(x,\mu^{2})T(b) \right) \bigg\vert_{r=r_s}  = \frac{1}{2}.
\end{equation}
This criterion equates to defining the saturation radius $r_{s}$ as the dipole size for which the proton consists of one interaction length.  Then the saturation scale is related to $r_{s}$ via $Q^{2}_{s}=2/r^{2}_{s}$. The impact parameter dependence of the IP-Sat framework is introduced through the proton thickness function which is taken to be a Gaussian:
\begin{equation}
T_{p}(b)=\frac{1}{2\pi B_{G}}e^{\frac{-b_{\perp}^2}{2B_{G}}},
\end{equation}
where $B_{G}=4.0$\,GeV$^{-2}$ is the average gluonic radius of the proton which follows from a fit to HERA diffractive data \cite{Rezaeian:2012ji}.  The gluon distribution function is initialized according to
\begin{equation}
xg(x,\mu_{0}^2)=A_{g}x^{-\lambda_{g}}(1-x)^{5.6}
\end{equation}
with $\lambda_{g}=0.058$ and $A_{g}=2.308$ taken from Ref.~\cite{Rezaeian:2012ji}, and $\mu_{0}^2 =1.0 ~{\rm GeV^2}$\footnote{In \cite{Schenke:2012fw}, $\mu_0^2 = 1.51$ GeV$^2$. This is the scale at which the gluon distribution is initialized, and it mostly effects the normalization of the energy density.}. This distribution function is then evolved using the leading order Dokshitzer-Gribov-Lipatov-Altarelli-Parisi (DGLAP)\cite{Dokshitzer:1977sg} \cite{Altarelli:1977zs}\cite{Gribov:1972ri} equations without quarks. The numerical solution to these equations appears in Ref.  \cite{DGLAP}. The scale relates to the dipole radius through
\begin{equation}
\mu^{2}=\frac{4}{r^{2}}+\mu_{0}^{2}
\end{equation}
The average color charge is of course zero, but there are local fluctuations of color charge that give rise to a non-zero expectation value for the average squared color charge density. This quantity is proportional to the saturation scale squared $Q_{s}^2 \propto g^{4}\mu^2$.  For the purposes of this paper, the constant of proportionality was determined by fitting the charged hadron multiplicity $dN_{ch}/d\eta$, a process which yielded $Q_{s} \approx 0.5 g^{2}\mu$\footnote{ $Q_s/(g^2\mu) \approx 0.75$ in \cite{Schenke:2012fw}}. The color charge configuration is then sampled from a Gaussian with variance 
\begin{equation}
\langle \rho_{A(B)}^{a}(\mathbf{x_{\perp}})\rho_{A(B)}^{b}(\mathbf{y_{\perp}}) \rangle=g^{2}\mu_{A(B)}^{2}(x,\mathbf{x_{\perp}})\delta^{ab}\delta^{2}(\mathbf{x_{\perp}}-\mathbf{y_{\perp}})
\end{equation}
where the subscripts $A$ and $B$ distinguish the two nuclei, where $A$ moves in the $+z$-direction and $B$ moves in the $-z$-direction.  Once sampled, this color charge distribution comprises the eikonal color current that sources the small-x classical gluon fields in the Color Glass Condensate framework,
\begin{equation}
J^{\nu}=\delta^{\nu\pm}\rho_{A(B)}(x^{\mp},\mathbf{x_{\perp}}).
\end{equation}
The gluon fields are then determined via the Classical Yang-Mills (CYM) equations
\begin{equation}
[D_{\mu},F^{\mu \nu}]=J^{\nu}.
\end{equation}
Working in covariant gauge and light cone coordinates prior to the collision, the CYM equations boil down to the Poisson equation,
\begin{equation}
A^{\pm}_{A(B)}=-\frac{\rho _{A(B)}}{\mathbf{\nabla_{\perp}}^{2}}.
\end{equation}
The pre-collision fields are pure gauge and can be transformed to light-cone gauge, which is more physical after the collision, via the path-ordered Wilson Line, 
\begin{equation}
V_{A(B)}(x_{\perp})=P\exp{ \left(-ig\int dx^{-}\frac{\rho^{A(B)}(x^{-},x_{\perp})}{\mathbf{\nabla_{\perp}}^{2}-m^{2}} \right)}
\end{equation}
which is discretized for the numerical simulation,
\begin{equation}
V_{A(B)}(x_{\perp})=\displaystyle\prod_{i=1}^{N_{y}} \exp{\left(-ig\frac{\rho_{i}^{A(B)}(x_{\perp})}{\nabla_{\perp}^{2}-m^{2}} \right)}
\end{equation}
where $N_{y}=10$, and $m=\lambda_{QCD}=0.2$ GeV is an infrared regulator that can be regarded as incorporating color confinement \cite{Lappi:2007ku}. Then the pure gauge fields prior to the collision can be written as
\begin{equation}
A^{i}_{A(B)}=\theta(x^{-(+)})\frac{i}{g}V_{A(B)}(\mathbf{x_{\perp}})\partial_{i}V_{A(B)}^{\dagger}(\mathbf{x_{\perp}}).
\end{equation}
Due to our choice of lightcone gauge, the other components vanish
\begin{equation}
A^{-(+)}=0.
\end{equation}
The gauge fields immediately after the collision can be found by matching the fields on the light cone \cite{PhysRevD.52.6231} \cite{PhysRevD.52.3809},
\begin{equation}
A^{i}=A_{A}^{i}+A_{B}^{i},
\end{equation}
\begin{equation}
A^{\eta}=\frac{ig}{2}[A_{A}^{i},A_{B}^{i}].
\end{equation}
The initial fields are then evolved using a lattice implementation of the sourceless CYM equations \cite{PhysRevC.67.054903}. After evolving to a matching time of $\tau_\mathrm{0}=0.4$\,fm, the stress energy tensor is constructed from the chromo-electric and chromo-magnetic fields. For higher numerical accuracy in the lattice spacing, an improved expression has been used for the square of the longitudinal component of the chromo-magnetic field $(B^{\eta})^2$ in the stress energy tensor, as compared to the expression used in equation 30 of \cite{Schenke:2012fw}. This improves the numerical stability in solving the eigenvalue problem
\begin{equation}
T^{\mu}_{\ \nu}u^{\nu}=\varepsilon u^{\mu},
\label{eq16}
\end{equation}
which yields the local energy density $\epsilon$ and flow velocity $u^{\mu}$ that are used to initialize the hydrodynamic evolution. 

In this work, the shear stress tensor $\pi^{\mu\nu}$ and bulk viscous pressure $\Pi$ are set to zero at the initial time of our hydrodynamic simulations. We note that setting $\pi^{\mu\nu}=\Pi=0$ introduces a discontinuity in matching the energy momentum current $T^{\tau \mu}$ from the Yang-Mills phase to the hydrodynamic phase at $\tau_0 = 0.4$ fm. An alternative way to avoid such a discontinuity is proposed in Ref. \cite{Gardim:2011qn}. But this procedure modifies the spatial component of the $T^{\mu\nu}$ tensor at the matching. 
A more proper way to avoid this discontinuity is to initialize hydrodynamic simulations with the values of $\pi^{\mu\nu}$ and $\Pi$ from the Glasma phase. The RMS values of these dissipative components of the stress energy component $\sqrt{\pi^{\mu\nu} \pi_{\mu\nu}}$ and $\vert \Pi \vert$ are about 15\% of the system's local enthalpy $\epsilon+P$ at $\tau_{0}=0.4$ fm. The effects of non-zero initial shear stress tensor on hadronic flow observables were studied in Ref. \cite{Vujanovic:2016anq}. The influence on particle spectra and elliptic flow was shown to be negligible for the conditions used in that work. Thus, we devote a more quantitative study on the effects of initial $\pi^{\mu\nu}$ and initial bulk pressure $\Pi$ with the IP-Glasma initial conditions to a future project.

\subsection{Centrality Selection}
In relativistic heavy-ion experiments, the collision centrality of an event is determined by its measured charged hadron multiplicity. However, fully simulating the necessary number of minimum bias collisions is time-consuming. The system's initial total energy in the mid-rapidity region, $dE/d\eta_s\vert_{\eta_s=0}$, is proportional to the final charged hadron multiplicity, which serves as a good approximation in numerical simulations. The IP-Glasma model is made up of classical gluon fields that can interact at finite impact parameter, and such interactions contribute to the system's energy-momentum tensor. This subtlety leads to some ambiguity in determining the threshold for whether a collision occurred in the numerical simulation, a notion that corresponds to setting the 100\% centrality boundary. In light of this, we sampled a large number of events in the impact parameter range from $0-20$ fm, and sorted the events by their total energy at mid-rapidity, $dE/d\eta_s\vert_{\eta_s = 0}$.  We then simulated a subset ($\sim$200 events per 10\% centrality bin) of the IP-Glasma initial conditions using our hybrid framework \cite{Ryu:2015vwa} to map the energy of IP-Glasma $dE/\tau d\eta_s\vert_{\tau=0.4}$, in units of GeV/fm, to the charged hadron multiplicity $dN_\mathrm{ch}/d\eta\vert_{\vert \eta \vert < 0.5}$. We fit the results with a power law curve to determine the following empirical relationship, 

\begin{equation} \label{eq:dNdy}
\frac{dN_\mathrm{ch}}{d\eta}\bigg\vert_{\vert \eta \vert < 0.5}=0.839\left( \frac{dE}{\tau d\eta_s}\bigg\vert_{\tau=0.4 \, {\rm fm}}  \right)^{0.833}.
\end{equation}


Using this formula, we varied the 100\% centrality cutoff until the ratios of multiplicities between centrality bins, say $dN_{ch}/d\eta$(0-5\%)/($dN_{ch}/d\eta$(5-10\%)),  matched the ratios from the experimental data. Once the 100\% boundary was determined, all other centralities became fixed. We then adjusted the overall normalization by a factor of $0.89$ to fix the magnitude of the energy density such that the measured $dN_\mathrm{ch}/d\eta\vert_{\vert \eta \vert < 0.5}$ in 0-5\% centrality was reproduced.

Our 100\% centrality cutoff yields a geometric cross section that is consistent with measured values for the total cross section of Pb-Pb collisions at 2.76 TeV \cite{Abelev:2013qoq}. In terms of energy, at 5.02 TeV the 100\% centrality cutoff corresponded to $dE/d\eta_s \vert_{\eta_s = 0}\approx 11 $\,GeV.  Taking the average particle transverse momentum to be 500 MeV, this amounts to the creation of $\sim22$ particles per unit rapidity. This is comparable to high multiplicity p+p events at the same energy \cite{Khachatryan:2016txc}, which are believed to constitute the lower bound for system size for hydrodynamics to be applicable. Thus our centrality cut is consistent with the hydrodynamics framework that we employ. To test the sensitivity of centrality selection on the choice of the cut-off energy, we varied the boundary by increasing the cut-off to $dE/d\eta_s \vert_{\eta_s = 0}\approx 22 $\,GeV. We found the charged hadron multiplicity in central 0-5\% centrality bin increased less than 1\%. The centrality dependence of various hadronic observables was insensitive to the precise choice of this cut-off energy.

\subsection{Hydrodynamics and Hadronic Cascade}
We evolved 1,500 IP-Glasma events per 10\% centrality bin assuming boost-invariance in the longitudinal direction. Starting at $\tau_\mathrm{0} = 0.4$ fm, every IP-Glasma initial condition was matched to hydrodynamic variables, namely the local energy density and flow velocity via Eq.(\ref{eq16}), and  evolved using the state-of-the-art viscous hydrodynamic model {\tt MUSIC} \cite{Schenke:2010rr}, together with the lattice QCD based equation of state (EoS), s95p-v1 \cite{Huovinen:2009yb}. Shear and bulk viscosities and their non-linear coupling terms are included in the hydrodynamic evolution \cite{Denicol:2014vaa}. It has previously been shown using Bayesian methodology that a non-zero temperature dependent bulk viscosity is favored by the data \cite{PhysRevC.94.024907}, and is necessary to describe transverse momentum and flow data simultaneously. In the hadronic phase, individual fluid cells are converted into particles on an isothermal hyper-surface, $T_\mathrm{sw} = 145$\,MeV, using the well-known Cooper-Frye procedure \cite{Shen:2014vra}. Each hydrodynamic freeze-out surface is oversampled for 100 events. Every sampled particle event is individually fed into a hadronic cascade model {\tt UrQMD} \cite{Bass:1998ca,Bleicher:1999xi} to simulate microscopically hadronic scatterings, baryon anti-baryon annihilations, and resonance decays in the dilute hadronic phase.
Particles are sampled in a boost-invariant fashion over 10 units of rapidity on the freeze-out surface. This ensures that the final particle sample after hadronic cascade remains boost invariant in the rapidity region $-2.5$ to $2.5$. The leaking of particles at large forward and backward rapidities does not affect the mid-rapidity observables. 

The switching temperature $T_\mathrm{sw}$ between the macroscopic hydrodynamic evolution and the microscopic transport description is fixed to reproduce the measured identified particle yields, especially the ratio between pion and proton yields (see Fig. \ref{fig4}c below). Because the bulk viscosity at the freeze-out affects the particle yields, it can lead to a change in the switching temperature compared to the chemical freeze-out temperature from the thermal model fit \cite{Floris:2014pta}. In addition, the baryon and anti-baryon annihilation in the hadronic transport phase changes the relative yields between mesons and baryons. It also affects the actual value of the switching temperature \cite{Steinheimer:2012rd}. A detailed study on the hadronic observables dependence on the switching temperature is shown in Ref. \cite{Ryu:2015vwa, Ryu:2017qzn}.
The choices of specific shear and bulk viscosities have large effects on the development of hydrodynamic flow (both radial and anisotropic flow) during the evolution. We choose an effective $\eta/s = 0.095$ and a temperature dependent specific bulk viscosity, $\zeta/s(T)$ in the simulations which provide a good description of charged hadron anisotropic flow coefficients and mean $p_T$ measurements.  The specific bulk viscosity is parameterized as follows,
\begin{widetext}
\begin{equation}
\zeta/s(T) = 0.9 \times \left\{
	\begin{array}{lcl}
		0.9e^{\left(\frac{T}{T_\mathrm{p}}-1\right)/0.0025} + 0.22e^{\left(\frac{T}{T_\mathrm{p}}-1\right)/0.022} + 0.03 & \mbox{for} & T < 0.95T_\mathrm{p} \\
		-13.77\left(\frac{T}{T_\mathrm{p}}\right)^2 + 27.55 \left(\frac{T}{T_\mathrm{p}}\right) - 13.45 & \mbox{for} & 0.95 T_\mathrm{p} < T < 1.05 T_\mathrm{p} \\
		0.025e^{-\left(\frac{T}{T_\mathrm{p}}-1\right)/0.025} + 0.25e^{-\left(\frac{T}{T_\mathrm{p}}-1\right)/0.13} + 0.001 & \mbox{for} & T > 1.05T_\mathrm{p} 
	\end{array}
			\right.,
\label{eq17}
\end{equation}
\end{widetext}
It is a parameterization based on Ref.~\cite{Karsch:2007jc,NoronhaHostler:2008ju}. The peak temperature of $\zeta/s(T)$ is set to $T_\mathrm{p} = 180$ MeV near the peak of the trace anomaly from the lattice calculation \cite{Bazavov:2014pvz}, where the conformal symmetry is maximally broken. In the current work, the same parametrization is used as in Ref.~\cite{Ryu:2015vwa}, except the overall normalization of $\zeta/s(T)$ is reduced by 10\% to account for some small variations $- O(10\%) - $ in the pre-equilibrium flow at $\tau_{sw} = 0.4$ fm.

After the hydrodynamic simulation, the spatial and momentum distribution of particles are sampled using the Cooper-Frye freeze-out procedure, 
\begin{equation}
E \frac{d N}{d^3 p}(x) = p^\mu d^3 \sigma_\mu(x) (f_\mathrm{eq}(x, p) + \delta f (x, p)),
\end{equation}
where $d^3 \sigma_\mu(x)$ is the normal vector on the freeze-out hypersurface and $f_\mathrm{eq}(x, p)$ is particle's thermal equilibrium distribution function in a freeze-out fluid cell with local temperature $T(x)$. The out-of-equilibrium corrections $\delta f(x, p)=\delta f^\mathrm{shear}(x, p) + \delta f^\mathrm{bulk}(x, p)$ are the first order shear and bulk viscous corrections to the thermal equilibrium distribution function. 
We used the same form of $\delta f^\mathrm{shear}$ and $\delta f^\mathrm{bulk}$ as in Ref. \cite{Ryu:2015vwa}, assuming the relaxation time approximation,
\begin{equation}
\delta f^\mathrm{shear} = f_\mathrm{eq}(1 \pm f_\mathrm{eq})\frac{\pi^{\mu\nu}p_\mu p_\nu}{2 T^2 (e + P)}
\label{eq.sheardeltaf}
\end{equation}
and
\begin{equation}
\delta f^\mathrm{bulk} = f_\mathrm{eq}(1 \pm f_\mathrm{eq})\left(\frac{-\Pi}{\zeta/\tau_\Pi}\right)\frac{1}{3T}\left(\frac{m^2}{E} - (1-3c_s^2) E\right).
\label{eq.bulkdeltaf}
\end{equation}
Here $\pi^{\mu\nu}$ is the shear stress tensor, $\Pi$ is the bulk pressure, and $\tau_\Pi$ is the relaxation time for bulk viscosity. The effects of $\delta f$ on hadronic flow observables will be discussed in the Appendix.

\subsection{Flow analysis with finite number of particles}

To compute hadronic flow observables, we first construct the flow vectors ${\bf Q}_n$ for each hydrodynamic event. The oversampled UrQMD events from the same hydrodynamic freeze-out surface are combined together into a super-event from which the ${\bf Q}_n$ vectors are computed,
\begin{equation}
{\bf Q}_n = \sum_{k=1}^{N^\mathrm{oversample}_\mathrm{ev}} \sum_{j=1}^{N^k_\mathrm{particle}} e^{i n \phi_j}.
\label{eq.Qn}
\end{equation}
Here the index $j$ runs over all particles in one UrQMD event with transverse momentum $p_T > 0.2$ GeV and the index $k$ runs over all the UrQMD events initialized with particle samples from the same hydrodynamic hyper-surface. Since the oversample factor $N^\mathrm{oversample}_\mathrm{ev} = 100$ in our calculations, the large particle multiplicity reduces the random fluctuation in ${\bf Q}_n$ vector that arise from sampling a finite number of particles. The error of the ${\bf Q}_n$ with respect to its value in the infinite number of particles limit can be estimated by calculating the event-plane resolution,
\begin{eqnarray}
R_n &=& \sqrt{\langle \cos (n (\Psi_n^A - \Psi_n^B)) \rangle_\mathrm{hydro\,ev}} \nonumber \\ 
&=&\sqrt{\left\langle \frac{{\bf Q}_n^A \cdot ({\bf Q}_n^B)^* }{\vert Q_n^A\vert \vert Q_n^B\vert} \right\rangle_\mathrm{hydro\,ev}}.
\end{eqnarray}
We choose the two subevents to be from the rapidity regions $-2.5$ to $-0.5$ and $0.5$ to $2.5$.
\begin{figure}[h!]
  \centering
    \includegraphics[width=0.95\linewidth]{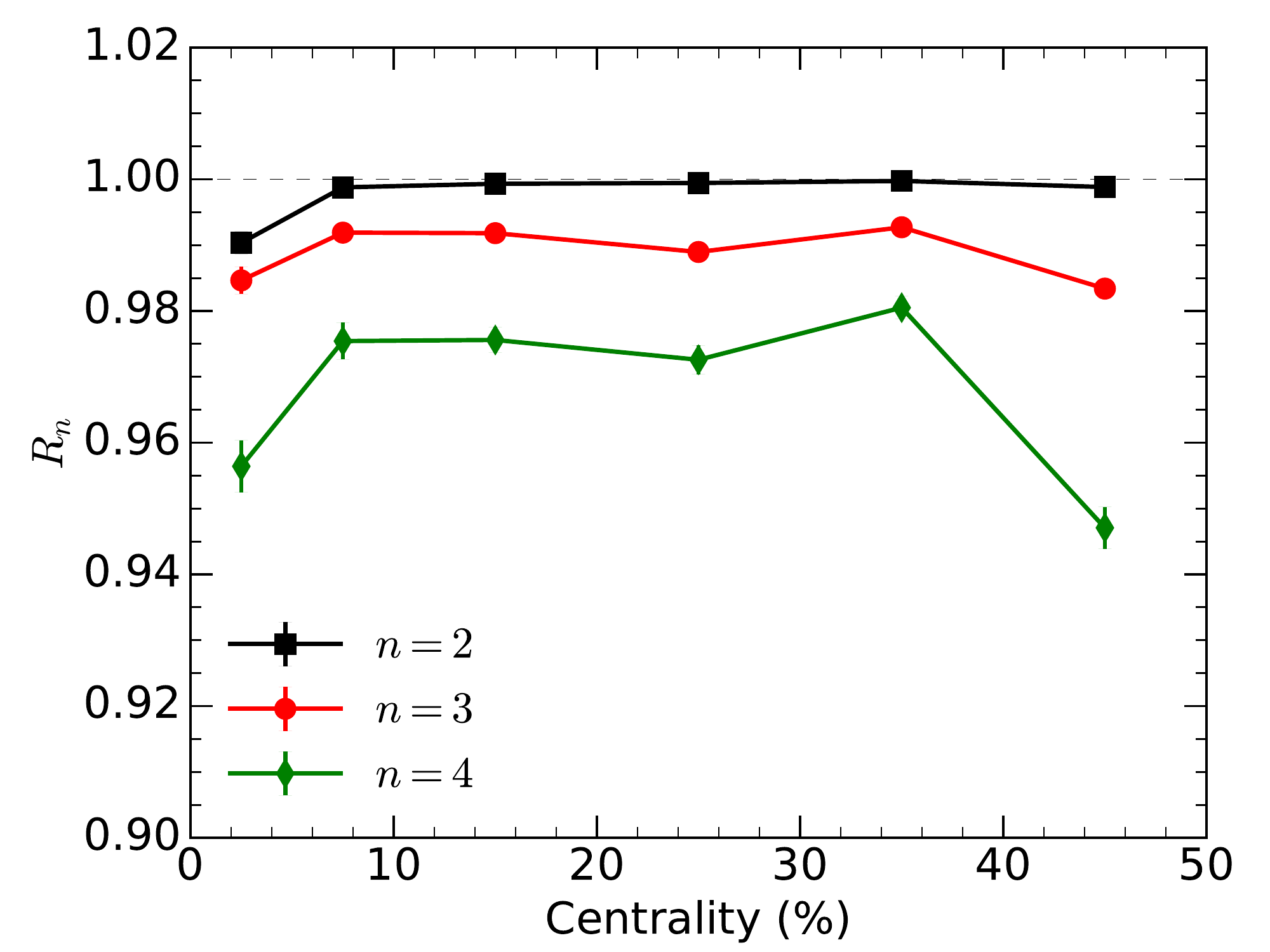} 
  \caption{(Color online) The centrality dependence of the event plane resolution factor $R_n$ for $n = 2, 3, 4$ in Pb+Pb collisions at 5.02 TeV.}
  \label{fig_reso}
\end{figure}
The values of the event-plane resolution factors is shown in Fig.~\ref{fig_reso} as a function of centrality for $n = 2, 3, 4$. For elliptic flow, the deviation of $R_2$ from 1 is less than 1\% from central to 50\% centrality. The resolution gets slightly worse for higher order $n$. For $v_4$, the $R_4$ stays within 5\% from unity. Compared to the typical values of $R_n$ in the experiments \cite{ATLAS:2012at}, our resolution factor is much higher. Such a high resolution of the flow vector ${\bf Q}_n$ in every hydrodynamic event ensures that our theoretical event-by-event $v_n$ distribution can be directly compared with the ATLAS measurements in which the finite resolution smearing was corrected using the Bayesian statistical unfolding method \cite{Aad:2013xma}.

Because there is no correlation between particles from different UrQMD events, the short range non-flow correlation, such as correlations from resonance decays, is suppressed by the oversampling factor $N^\mathrm{oversample}_\mathrm{ev}$ when we compute the multi-particle correlation functions using the ${\bf Q}_n$ vectors.
The two-particle cumulant flow coefficients are computed as,
\begin{equation}
v_n\{2\} = \frac{\langle {\rm Re} \{{\bf Q}_n \cdot ({\bf Q}_n)^*\}  \rangle_\mathrm{hydro\,ev}}{\langle N^2 \rangle_\mathrm{hydro\,ev}}
\end{equation}
and the $p_T$-differential flow coefficients from the scalar product method is, 
\begin{equation}
v_n\{\mathrm{SP}\}(p_T) = \frac{\langle {\rm Re}\{{\bf Q}_n(p_T) \cdot ({\bf Q}^\mathrm{ref}_n)^*\} \rangle_\mathrm{hydro\,ev}}{\langle N(p_T) N^\mathrm{ref} \rangle_\mathrm{hydro\,ev} v_n\{2\}}.
\end{equation}
Here ${\rm Re} \{\cdots\}$ takes the real part of the correlation function and $N^\mathrm{ref}$ and $N(p_T)$ are the corresponding particle multiplicities for ${\bf Q}^\mathrm{ref}_n$ and ${\bf Q}(p_T)$ vectors, respectively.

\section{Results and Discussion}

In this section, we will start by comparing the numerical model to existing experimental measurements of Pb+Pb collisions at 2.76 TeV. By regarding the good descriptions of the existing experimental data as a base point, we extrapolate our calculation to higher collision energy and make predictions and postdiction for various hadronic flow observables.

\subsection{Charged hadron yields and anisotropic flow}

\begin{figure}[h!]
  \centering
    \includegraphics[width=0.95\linewidth]{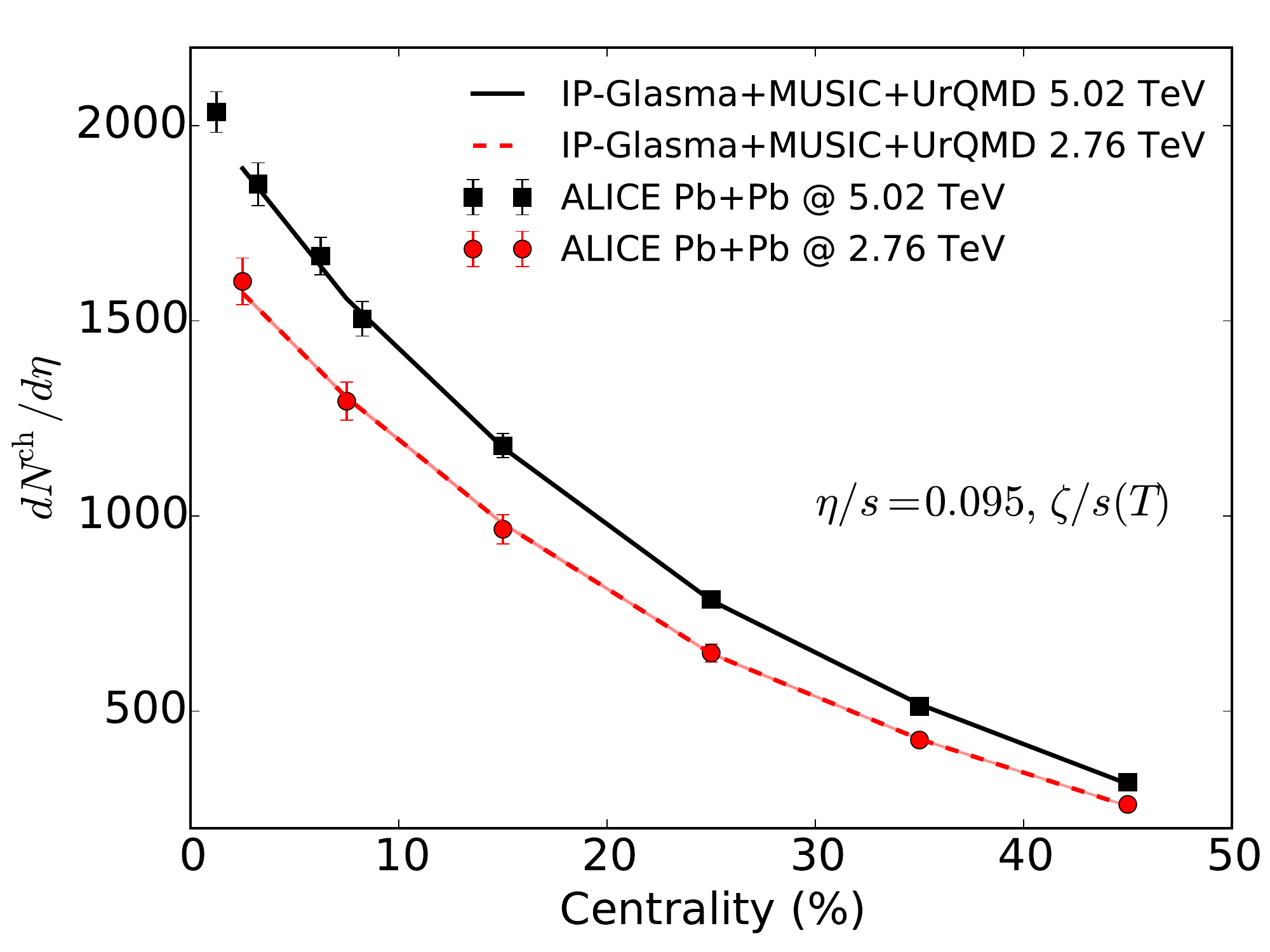} 
  \caption{(Color online) The centrality dependence of charged hadron multiplicity in Pb+Pb collisions compared with the ALICE measurements at 2.76 TeV \cite{Aamodt:2010cz} and 5.02 TeV \cite{Adam:2015ptt}.}
  \label{fig1}
\end{figure}

\begin{figure*}[ht!]
  \centering
  \begin{tabular}{cc}
  \includegraphics[width=0.45\linewidth]{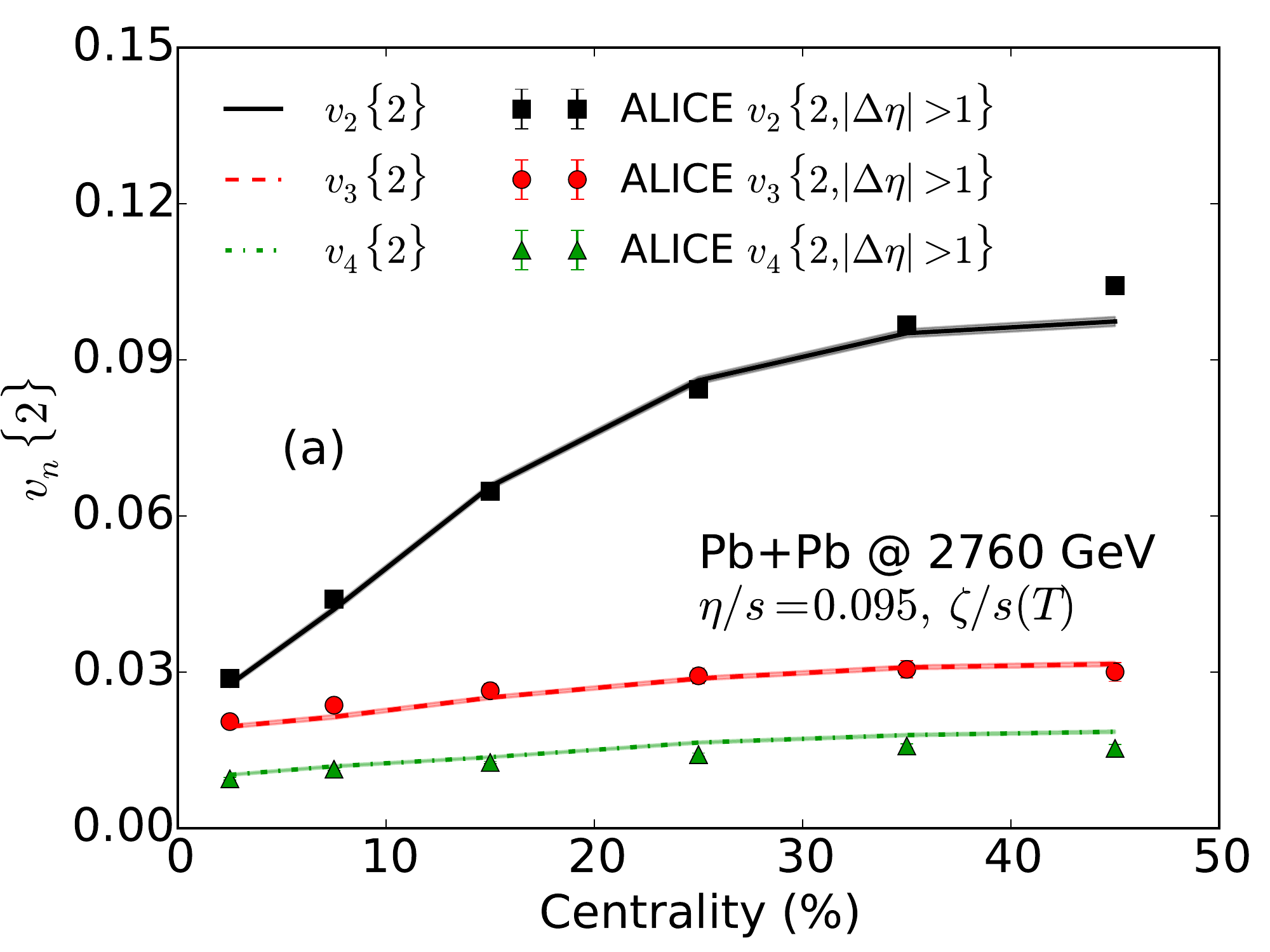} &
  \includegraphics[width=0.45\linewidth]{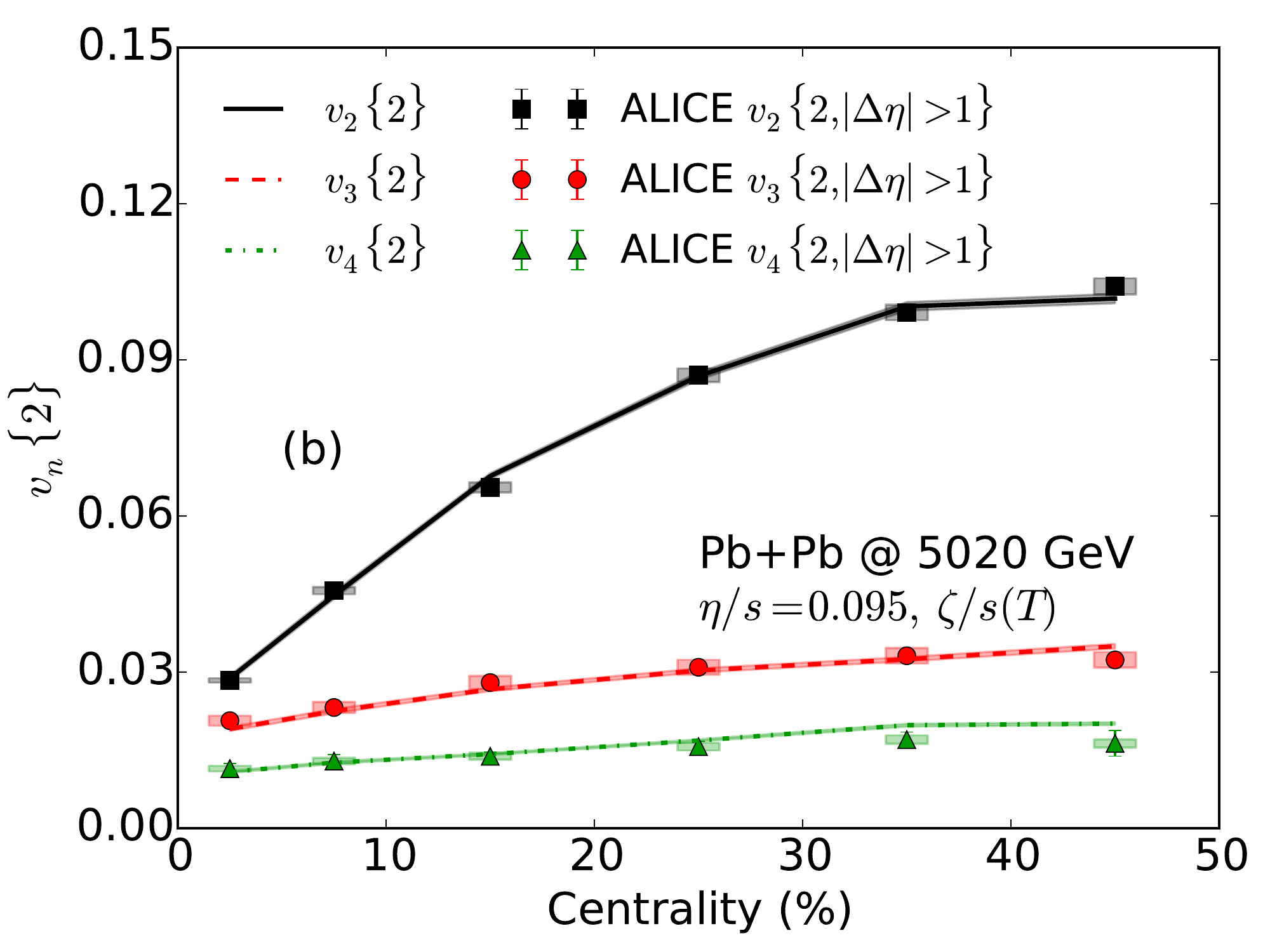}
  \end{tabular}
  \caption{(Color online) The centrality dependence of charged hadron anisotropic flow coefficients in Pb+Pb collisions at 2.76 TeV $(a)$ and 5.02 TeV $(b)$. The integrated $p_T$ range for the two particle cumulant $v_n\{2\}$ is from 0.2 to 3.0 GeV. Theoretical results are compared with the recent ALICE measurements at both collision energies \cite{ALICE:2011ab,Adam:2016izf}.}
  \label{fig2}
\end{figure*}
%

To calibrate our hybrid simulations to the desired collision energies, we adjust an overall normalization factor on the system's energy density such that the final charged hadron multiplicity agrees with the experimental measurements. In Fig.~\ref{fig1}, charged hadron multiplicities at mid-rapidity are shown as a function of collision centrality for Pb+Pb collisions at 2.76 TeV and 5.02 TeV. We used the same overall normalization $\sim0.89$ on the initial energy density to reproduce the measured $dN^\mathrm{ch}/d\eta \vert_{\vert \eta \vert < 0.5}$ in 0-5\% most central collisions at both collision energies. The centrality dependence of the charged hadron multiplicity is well described by the IP-Glasma model. The viscous entropy production during the hydrodynamic evolution is found to be about $\mathcal{O}(10\%)$ compared to the initial entropy in the system.

Charged hadron anisotropic flow coefficients are shown as a function of collision centrality in Figs.~\ref{fig2}. With $\eta/s = 0.095$ and $\zeta/s(T)$ in Eq.~(\ref{eq17}), our hybrid simulations can reproduce ALICE measurements for $v_2\{2\}$, $v_3\{2\}$, and $v_4\{2\}$ up to 40-50\% semi-peripheral collisions at 2.76 TeV quite well, as shown in Fig.~\ref{fig2}a. The simultaneous description  to all the $v_n$ coefficients as well as to their event-by-event distribution (see Fig.~\ref{fig7} below) suggests that our new formulation of the IP-Glasma model is consistent with previous works \cite{Schenke:2012fw, Schenke:2012wb,Ryu:2015vwa}. Now, by keeping the same values for the transport coefficients in hydrodynamic simulations, our $v_{n}\{2\}$ results in Pb+Pb collisions at 5.02 TeV agree with the latest ALICE data within small experimental error bars \cite{Adam:2016izf}. In our simulations, the charged hadron $v_2\{2\}$ increased by (4.1 $\pm$ 1.7)\%, $v_3\{2\}$ by (5.1 $\pm$ 2.2)\%, and $v_4\{2\}$ by (6.2 $\pm$ 2.3) \% from 2.76 to 5.02 TeV. The relative increase of $v_n\{2\}$ agrees with the experimental values reported in Ref. \cite{Adam:2016izf}. The initial eccentricities $\varepsilon_n$ do not change at the two collision energies in the IP-Glasma model. The increase of $v_n$ is due to the longer fireball lifetime at the higher collision energy, which converts more initial spatial eccentricity into final particle momentum anisotropy. The fact that the same effective $\eta/s = 0.095$ can quantitatively reproduce the $p_T$-integrated charged hadrons $v_n$ at the two collision energies suggests the temperature ranges probed by the fireball are quite similar. At the initial time $\tau_0 = 0.4$ fm, the peak temperature of the system can reach to $\sim$450 MeV in central Pb+Pb collisions at 2.76 TeV. The $\sim$80\% increase in the collision energy results in a peak temperature that is about 20 MeV higher at 5.02 TeV.

\begin{figure*}[ht!]
  \centering
  \begin{tabular}{cc}
  \includegraphics[width=0.45\linewidth]{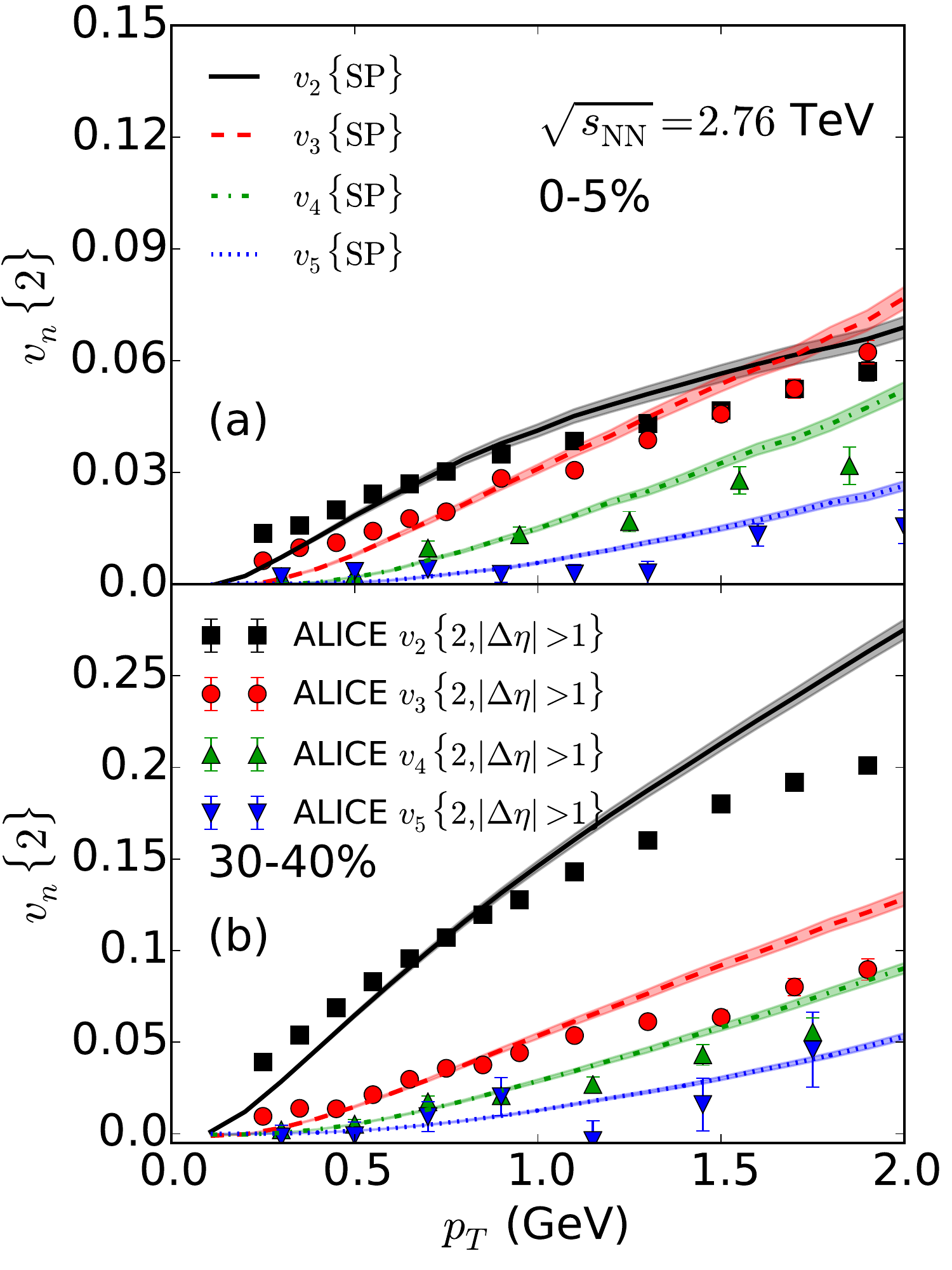} &
  \includegraphics[width=0.45\linewidth]{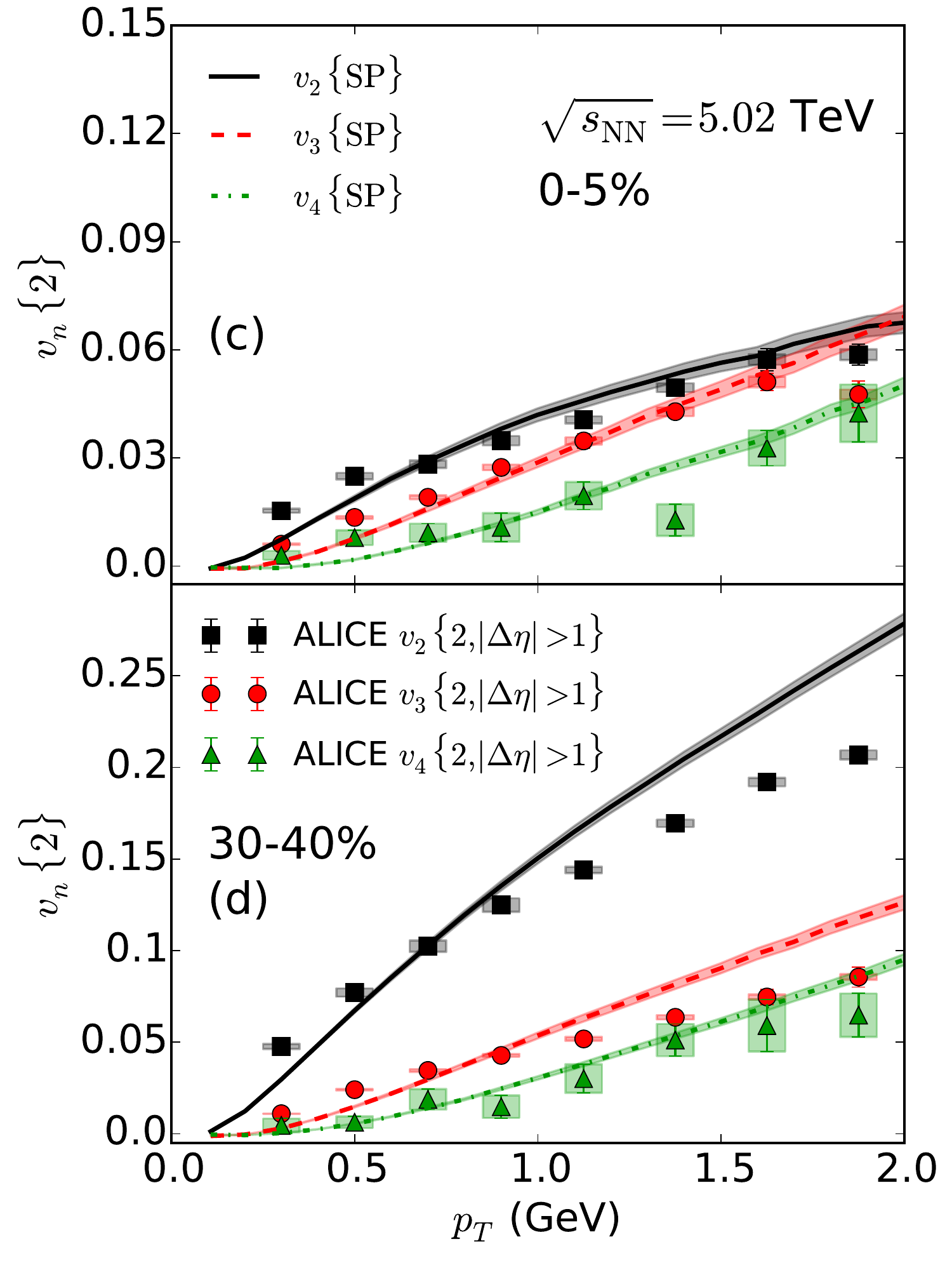}
  \end{tabular}
  \caption{(Color online) Charged hadron $p_T$-differential $v_n$ in 0-5\% $(a, c)$ and 30-40\% $(b, d)$ centrality bins in Pb+Pb collisions at 2.76 $(a, b)$ and 5.02 $(c, d)$ TeV. The reference flow vector is the charged hadron $v_n$ integrated from 0.2 to 3.0 GeV. At both collision energies, theoretical results are compared with the ALICE measurements \cite{ALICE:2011ab,Adam:2016izf}. }
  \label{fig3}
\end{figure*}
%

In Fig.~\ref{fig3}, charged hadron $p_T$-differential $v_n(p_T)$ are computed using the scalar-product method \cite{Luzum:2012da} and compared with the ALICE measurements at the two collision energies \cite{ALICE:2011ab,Adam:2016izf}. In the top 0-5\% centrality, our hybrid approach describes the ALICE data within 10\% up to 2 GeV for all the $v_n$ coefficients.
The agreement becomes limited to lower $p_T$ in 30-40\% peripheral collisions. Charged hadron $v_n$ at high $p_T$ are overestimated by 30\% at $p_T = 2$ GeV. However, we should note that particles $p_T$ spectra are underestimated for $p_T > 1.5$ GeV in 30-40\% centrality (see Figs.~\ref{fig5} below). In this intermediate $p_T$ region, one would expect that the contribution from jet shower and mini-jets begins to play an important role. The inclusion of these components will affect the charged hadron $v_n(p_T)$ in this intermediate $p_T$ range \cite{Andrade:2014swa,Schulc:2014jma,Crkovska:2016flo,Ryu:2016xxx}. The quality of the model for data descriptions is similar at the two collision energies. Compared to the Ref.~\cite{Gale:2012rq}, the agreement with the measured $v_n(p_T)$ is worse. This difference is due to the inclusion of bulk viscosity. The bulk viscosity plays a critical role in reducing the hydrodynamic radial flow and improves the description of the experimental measured mean $p_T$ \cite{Ryu:2015vwa}. At the same time, the inclusion of bulk viscosity requires a reduction in shear viscosity in order to describe the integrated $v_n$. Including the bulk viscosity and reducing the shear viscosity alter the shape of the $p_T$ dependence of the differential flow harmonics. The effect of out-of-equilibrium correction $\delta f$ on $p_T$-differential $v_n$ is investigated in the Appendix.

\subsection{Identified particle observables}

While charged hadron $v_n$ reflects a detailed flow pattern of the evolving medium at final kinetic freeze-out, identified particle observables, especially heavy baryons, are more sensitive to the background hydrodynamic flow. Furthermore, because the multi-strange hadrons, such as $\phi$, $\Lambda$, $\Xi^-$, and $\Omega$, have small scattering cross sections in the hadronic phase, their observables can shed light on how the medium flows at an earlier time prior to the kinetic freeze-out of the other light hadrons. 
\begin{figure*}[ht!]
  \centering
  \begin{tabular}{cc}
   \includegraphics[width=0.555\linewidth]{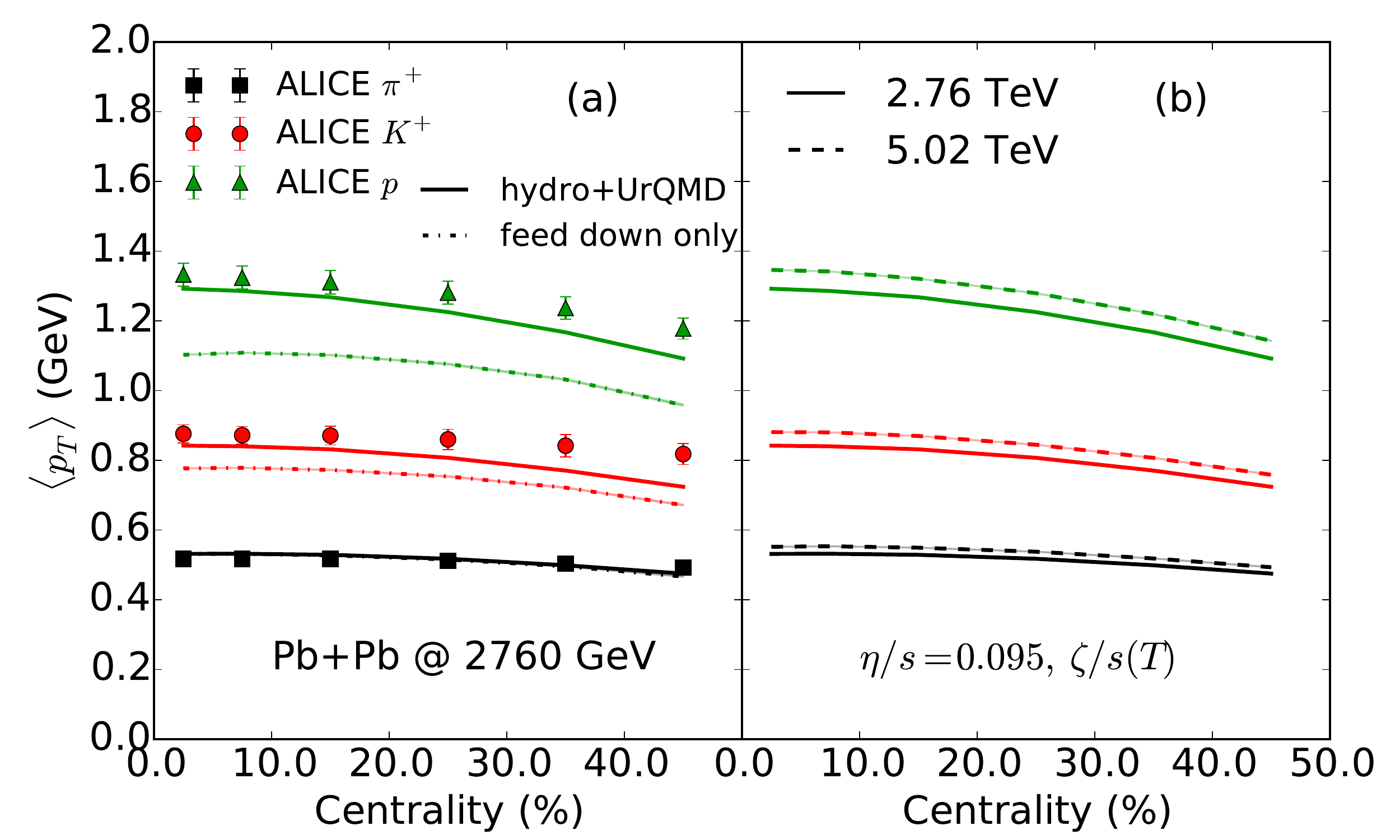}&
   \includegraphics[width=0.445\linewidth]{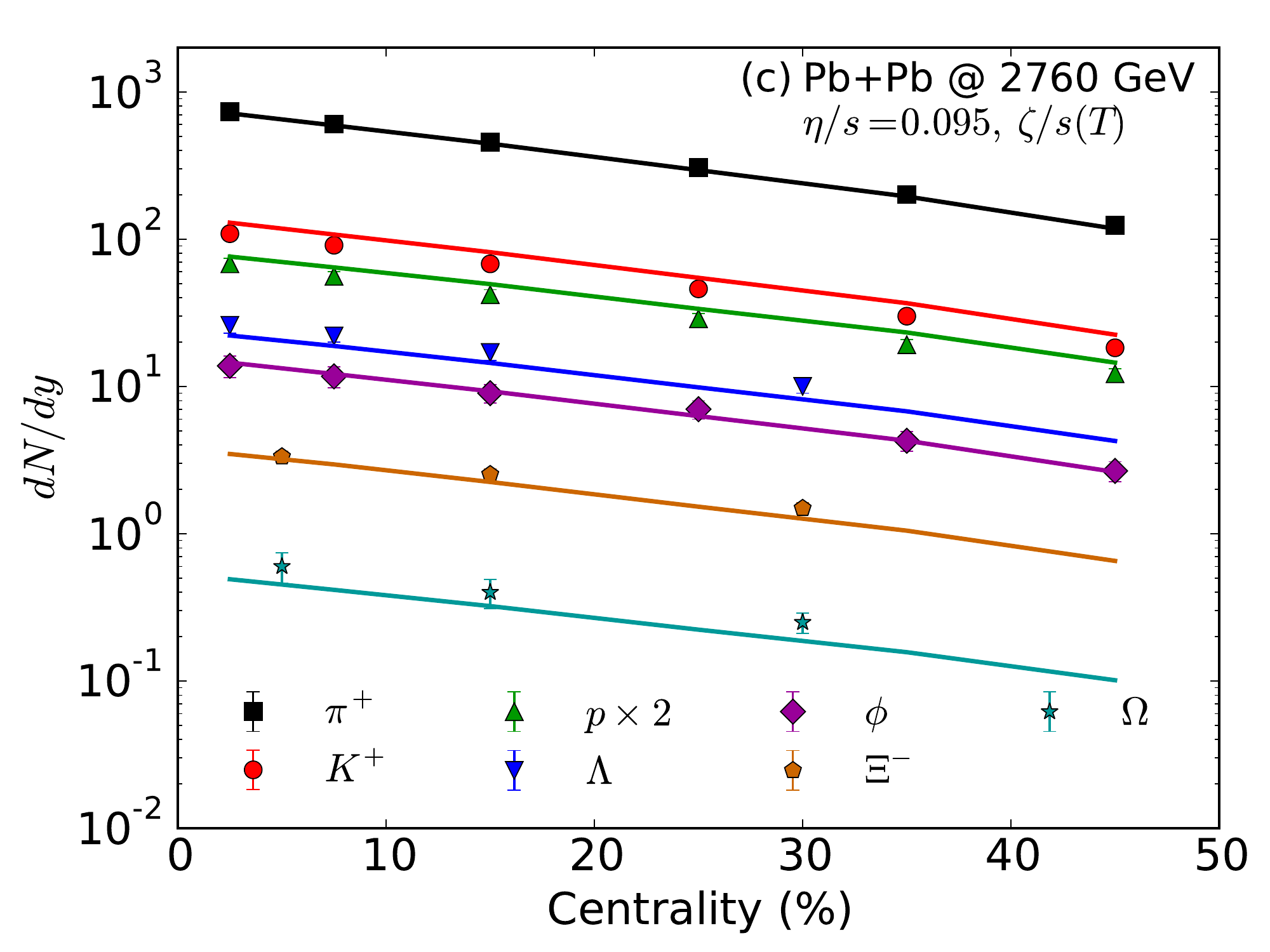}
  \end{tabular}
  \caption{(Color online) {\it Panel (a):} Identified particle averaged momentum, $\langle p_T \rangle$, as functions of centrality in Pb+Pb collisions at 2.76 TeV compared with the ALICE data \cite{Abelev:2013vea}.  {\it Panel (b):} Predictions of $\pi^+$, $K^+$, and $p$ mean $p_T$ in Pb+Pb collisions at 5.02 TeV. {\it Panel (c):} Identified particle yields as functions of centrality compared to the ALICE data.}
  \label{fig4}
  \end{figure*}

Hydrodynamic radial flow blueshifts individual hadrons to higher $p_T$, which increases their average transverse momentum. In Fig.~\ref{fig4}a, we compare light hadron mean-$p_T$'s as a function of centrality with the ALICE measurements \cite{Abelev:2013vea}. The $\langle p_T \rangle$ of $\pi^+$ agrees with experimental measurements through all centralities, while the proton mean $p_T$ in the semi-peripheral centralities are underestimated by $\sim10\%$. The dash-dotted lines represent results in which hadronic rescatterings were turned off. Although the light $\pi^+$ mean-$p_T$ is barely affected, hadronic scatterings increase the proton $\langle p_T \rangle$ by 30\%. This reflects that heavier protons show a larger sensitivity of hydrodynamic radial flow compared to the lightest $\pi$. In Fig.~\ref{fig4}b, we provide predictions for $\pi^+$, $K^+$, and proton mean-$p_T$ at 5.02 TeV. Compared to the results at 2.76 TeV, the mean $p_T$ of $\pi^+$, $K^+$, and $p$ increase by 4$\sim$5\%.
In Fig.~\ref{fig4}c, we compared the identified particle yields with the ALICE measurements in Pb+Pb collisions at 2.76 TeV. Our calculation agrees with the experimental data within the statistical errors. The centrality dependence is reproduced. We noted that the yields of strange hadrons, $\phi$, $\Xi$ and $\Omega$, are well described. Understanding the production of strange hadrons is an important topic for lattice QCD calculations \cite{Bellwied:2013cta,Bazavov:2014xya}. 

\begin{figure}[ht!]
  \centering
  \includegraphics[width=1.\linewidth]{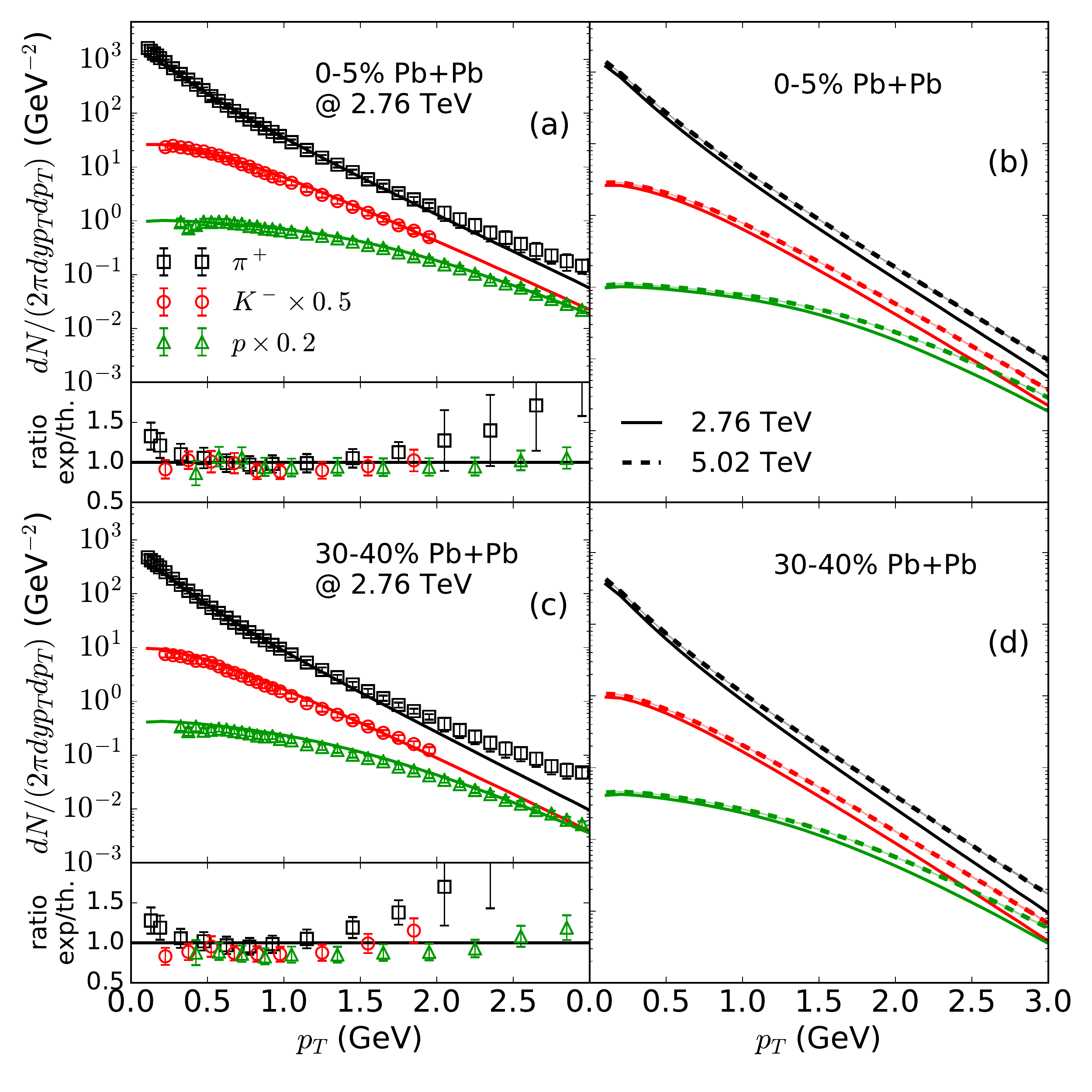}
  \caption{(Color online) {\it Panel (a, c):} Light hadrons $p_T$-differential spectra in 0-5\% and 30-40\% Pb+Pb collisions at 2.76 TeV compared with the ALICE measurements \cite{Abelev:2013vea}. {\it Panel (b, d):} Predictions of $\pi^+$, $K^+$, and $p$ spectra in Pb+Pb collisions at 5.02 TeV. }
  \label{fig5}
\end{figure}
%

Figs.~\ref{fig5}a,c present comparisons of $\pi^+$, $K^+$, and proton spectra with the ALICE data \cite{Abelev:2013vea} in 0-5\% and 30-40\% centralities Pb+Pb collisions at 2.76 TeV. Consistent with the mean $p_T$ results shown in  Fig.~\ref{fig4}a, our hybrid model can reproduce the $\pi^+$ spectra up to 2 GeV in 0-5\% centrality. The deviation from the data starts at $p_T\sim1.5$ GeV, in 30-40\% centrality bin. Again, contribution from jet showers and mini-jets is expected to be important in this intermediate $p_T$ range \cite{Ryu:2016xxx}. The agreement with the heavier proton spectra extends to higher $p_T$. This is mainly because heavier protons receive a stronger blueshift effect from the hydrodynamic radial flow. The predictions of identified particle spectra are presented in Figs.~\ref{fig5}b,d. Besides the increase in particle yields, the identified particle spectra are flatter reflecting a stronger radial flow developed at the higher collision energy.  

\begin{figure}[ht!]
  \centering
  \includegraphics[width=1.\linewidth]{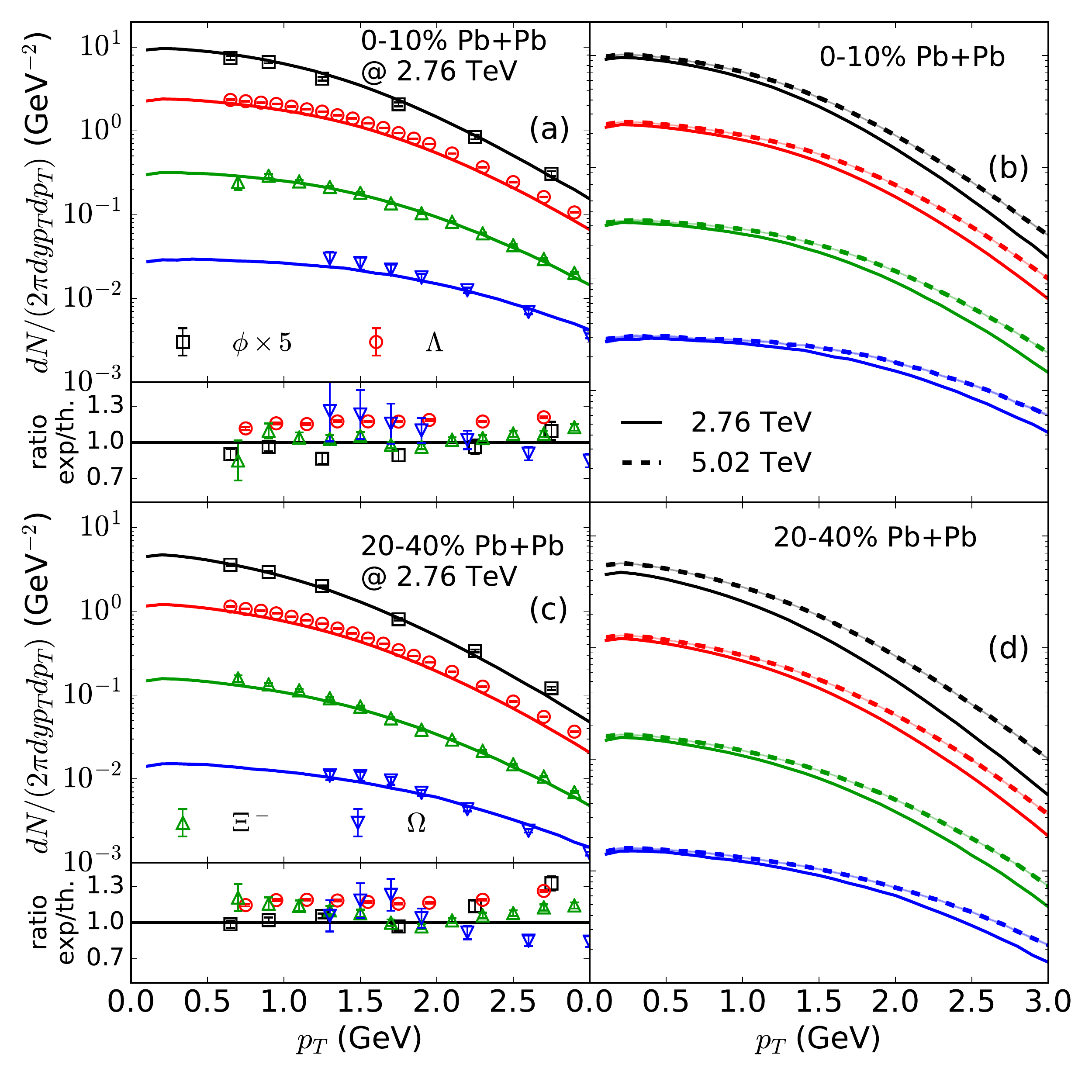}
  \caption{(Color online) {\it Panel (a, c):} Multi-strange hadrons $p_T$-differential spectra in 0-10\% and 20-40\% Pb+Pb collisions at 2.76 TeV compared with the ALICE measurements \cite{ABELEV:2013zaa, Abelev:2013xaa, Abelev:2014uua}. {\it Panel (b, d):} Predictions of $\phi$, $\Lambda$, $\Xi^-$, and $\Omega$ spectra in Pb+Pb collisions at 5.02 TeV. }
  \label{fig5b}
\end{figure}
%

The spectra of multi-strange particles are shown in Figs.~\ref{fig5b}. Since $\phi$ mesons are unstable, we need to reconstruct them from their decay daughters. We use $K^+K^-$ pairs to reconstruct $\phi$ mesons as described in the experimental analysis of Ref.~\cite{Abelev:2014uua}. In the {\tt UrQMD} results, in addition to checking that the invariant mass of $K^+K^-$ pairs is at the $\phi$ resonance peak, $M = 1.019 \pm 0.00433$ GeV \cite{Beringer:1900zz}, we require the last interactions of both $K^+$ and $K^-$ to be from decay processes and to be from the same space-time positions. Using these criteria, we can avoid the sophisticated statistical background subtraction performed in the experimental analysis \cite{Abelev:2014uua}. Finally, we correct the absolute yield of $\phi$ meson by its decay branching ratio to $K^+K^-$ pairs, $B_r(\phi \rightarrow K^+ K^-) = 0.489$ \cite{Beringer:1900zz}. Please note that in our reconstruction method, if a daughter kaon from a decay scatters with other particles in the hadronic phase, its mother $\phi$ meson will not be reconstructed, as in the experimental reconstruction procedure. Thus, the yield of reconstructed $\phi$ can be directly compared to the measurements in Ref.~\cite{Abelev:2014uua}. Figs.~\ref{fig5b}a,c show that our reconstructed $\phi$ meson samples provide an excellent description of the ALICE measured spectra up to 3 GeV \cite{Abelev:2014uua}. Both the absolute yield and the shape of $\phi$ meson spectra are well reproduced.

In the strange baryon sector, the $\Lambda$ measurement from ALICE \cite{Abelev:2013xaa} receives feed down contributions from $\Sigma^0$ baryons. This weak decay channel $\Sigma^0 \rightarrow \Lambda + \gamma$ is not included in the {\tt UrQMD} simulations. Thus, we perform two-body decays of $\Sigma^0$ to $\Lambda$ as an afterburner in our simulations after the system is kinetically frozen-out. Since the lifetime of $\Sigma^0$ is about 2000 fm/$c$, such decay processes will happen outside the fireball. This feed down contribution is about 30\% compared to the thermally produced $\Lambda$ yield. Figs.~\ref{fig5b}a,c shows our spectra of $\Lambda$, $\Xi^-$ and $\Omega$ compared with the ALICE measurements. We underestimated the measured $\Lambda$ spectra by $\sim$15\% in both central and semi-peripheral centrality bins. The shapes of spectra are well reproduced. The spectra of multi-strange baryons, $\Xi^-$ and $\Omega$, show good agreements with the ALICE data \cite{ABELEV:2013zaa}. 
Predictions of strange hadrons spectra in Pb+Pb collisions at 5.02 TeV are presented in Fig.~\ref{fig5b}b,d. Similar to the light hadron spectra, the stronger hydrodynamic radial flow at higher collision energy produces flatter spectra for multi-strange particles. 

\begin{figure}[ht!]
  \centering
  \includegraphics[width=1.\linewidth]{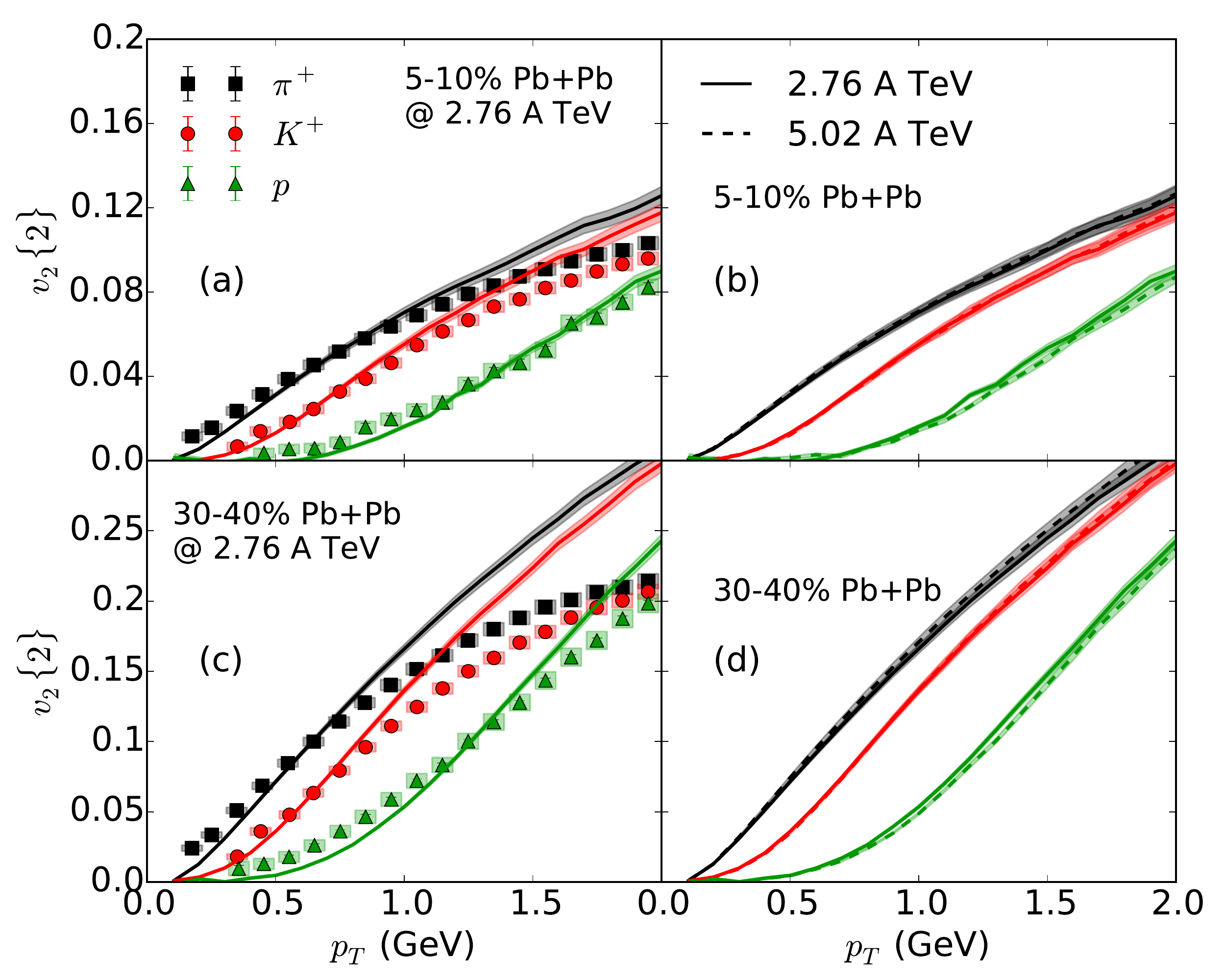}
  \caption{(Color online) {\it Panel (a, c):} The elliptic flow coefficients $v_2$ of identified particles in 5-10\% and 30-40\% Pb+Pb collisions at 2.76 TeV compared with the ALICE measurements \cite{Abelev:2014pua}. The reference flow vector is the charged hadron $v_2$ integrated from 0.2 to 3.0 GeV. {\it Panel (b, d):} Predictions of the $v_2(p_T)$ for $\pi^+$, $K^+$, and $p$ in Pb+Pb collisions at 5.02 TeV.}
  \label{fig6a}
\end{figure}
%

\begin{figure}[ht!]
  \centering
  \includegraphics[width=1.\linewidth]{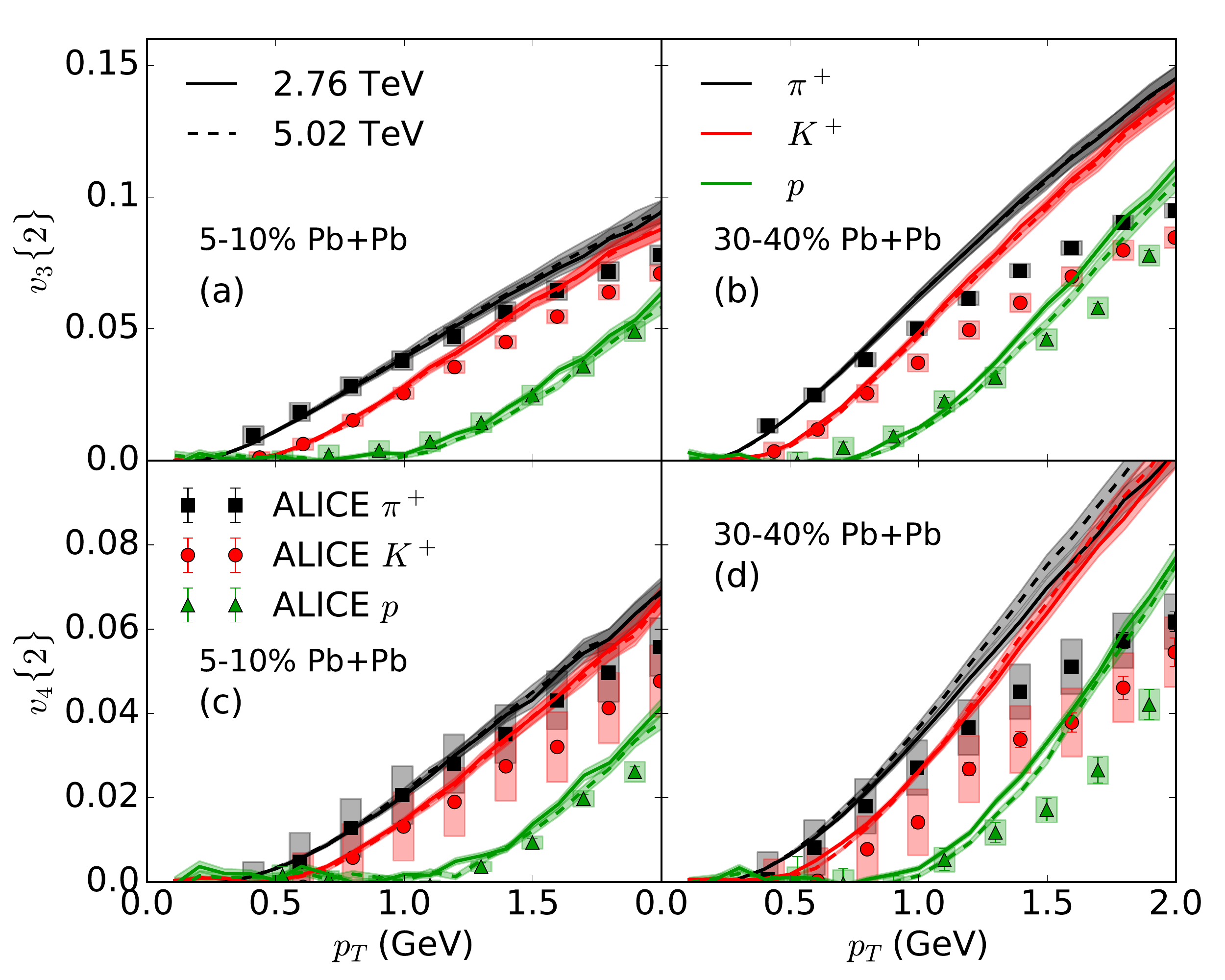}
  \caption{(Color online) Higher order anisotropic flow coefficients, $v_{3,4}\{2\}(p_T)$, for $\pi^+$, $K^+$, and $p$ in Pb+Pb collisions at 2.76 and 5.02 TeV. The reference flow vector is the charged hadron $v_n$ integrated from 0.2 to 3.0 GeV. The ALICE measurements at 2.76 TeV \cite{Adam:2016nfo} are shown for comparison.}
  \label{fig6b}
\end{figure}

The $p_T$-differential $v_n$ of identified hadrons can tell us how the momentum anisotropy of the system is distributed over different species of particles as a function of their transverse momentum. Figs.~\ref{fig6a}a,c show comparisons for identified hadron $v_2(p_T)$ between our hybrid approach and the ALICE measurements at 2.76 TeV. The trend of agreement for pion $v^{\pi^+}_2(p_T)$ is similar compared to the situation for the charged hadron $v_2(p_T)$ shown in Fig.~\ref{fig3}. The agreement with data extends to higher $p_T$ for heavier protons. This is mainly because the stronger blueshift from radial flow improves the model description of the proton spectra as shown in Fig.~\ref{fig5}, which helps the system to distribute proton's momentum anisotropy properly into different $p_T$ bins. Predictions at 5.02 TeV are presented in Figs.~\ref{fig6a}b,d. Compared to the $v_2(p_T)$ at 2.76 TeV, the variation at higher collision energy is very small. This is a consequence of the cancelling effects between the increase of $p_T$-integrated $v_2$ and stronger radial flow \cite{Shen:2012vn}. The former increases the differential $v_2$ at a fixed $p_T$ while the latter blueshifts the particles carrying large momentum anisotropy to higher $p_T$ region. The $v_2(p_T)$ of light hadrons remains unchanged. For proton $v_2(p_T)$, one can see that the blueshift effect slightly wins over the increase of elliptic flow. The proton $v_2(p_T)$ at higher collision energy is blueshifted to higher $p_T$ \cite{Shen:2011eg}. Higher order harmonic flow coefficients, $v_3$ and $v_4$, of identified particles are shown in Fig.~\ref{fig6b}. Comparisons with the ALICE measurements \cite{Adam:2016nfo} show similar quality as the $v_2$ comparison in Fig.~\ref{fig6a}. Compared to elliptic flow, the mass splitting between $\pi^+$ and proton are larger for higher order of $n$ \cite{Xu:2016hmp}. Predictions at 5.02 TeV are presented as dashed curves. The collision energy dependence of identified particle $v_{3,4}(p_T)$ is similar to those of elliptic flow. 

%
\begin{figure}[ht!]
  \centering
  \includegraphics[width=1.\linewidth]{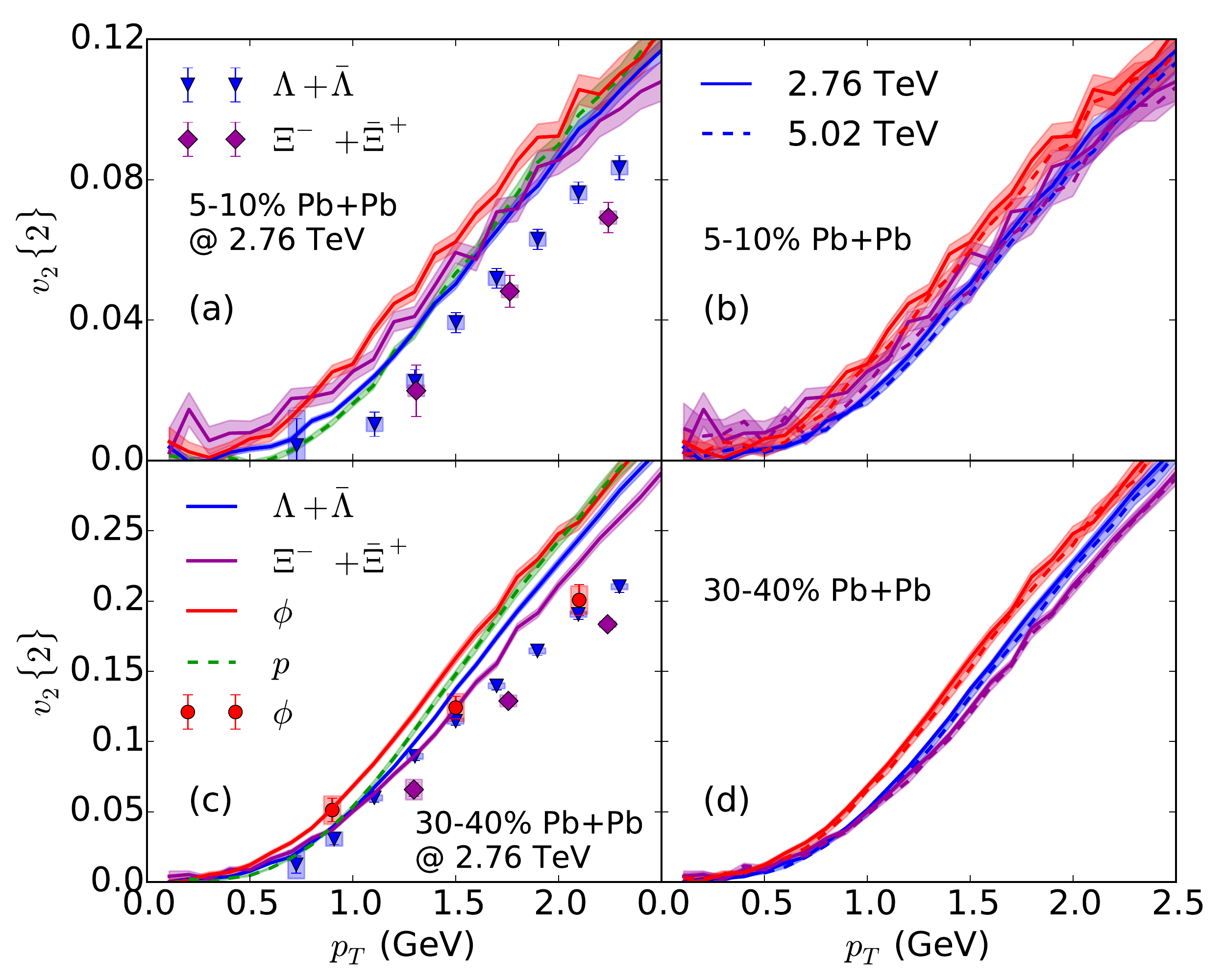}
  \caption{(Color online) {\it Panel (a, c):} The elliptic flow coefficients $v_2$ of multi-strange hadrons $\phi$, $\Lambda+\bar{\Lambda}$, and $\Xi^-+\bar{\Xi}^+$ in 5-10\% and 30-40\% Pb+Pb collisions at 2.76 TeV compared with the ALICE measurements \cite{Abelev:2014pua}. The reference flow vector is the charged hadron $v_2$ integrated from 0.2 to 3.0 GeV. {\it Panel (b, d):} Predictions of the $v_2(p_T)$ for $\phi$, $\Lambda+\bar{\Lambda}$, and $\Xi^-+\bar{\Xi}^+$ in Pb+Pb collisions at 5.02 TeV.}
  \label{fig6c}
\end{figure}
%

Fig.~\ref{fig6c} shows the elliptic flow coefficients of $\phi$, $\Lambda+\bar{\Lambda}$, and $\Xi^-+\bar{\Xi}^+$ from our hybrid calculations. Compared with the ALICE measurements at 2.76 TeV, the multi-strange hadron $v_2(p_T)$ are less blueshifted.
Because of the small scattering cross sections in the hadron transport phase, these multi-strange hadrons do not pick up enough blueshift from the ``pion wind''. The mass ordering of $v_2(p_T)$ of multi-strange hadrons is violated compared to proton $v_2(p_T)$.
A similar trend was found in other hybrid calculations \cite{Zhu:2015dfa,Takeuchi:2015ana,Takeuchi:2016czw, Zhu:2016qiv}. The same problem exists at both collision energies. 

\subsection{Flow distribution and correlations}

With model parameters being fixed by hadronic spectra and flow measurements, the following correlation observables are ``parameter-free'' predictions from our framework. These observables can further test the fluctuation spectrum of the initial state as well as the hydrodynamic response from which flow correlations are generated. 

\begin{figure}[ht!]
  \centering
  \begin{tabular}{c}
  \includegraphics[width=0.9\linewidth]{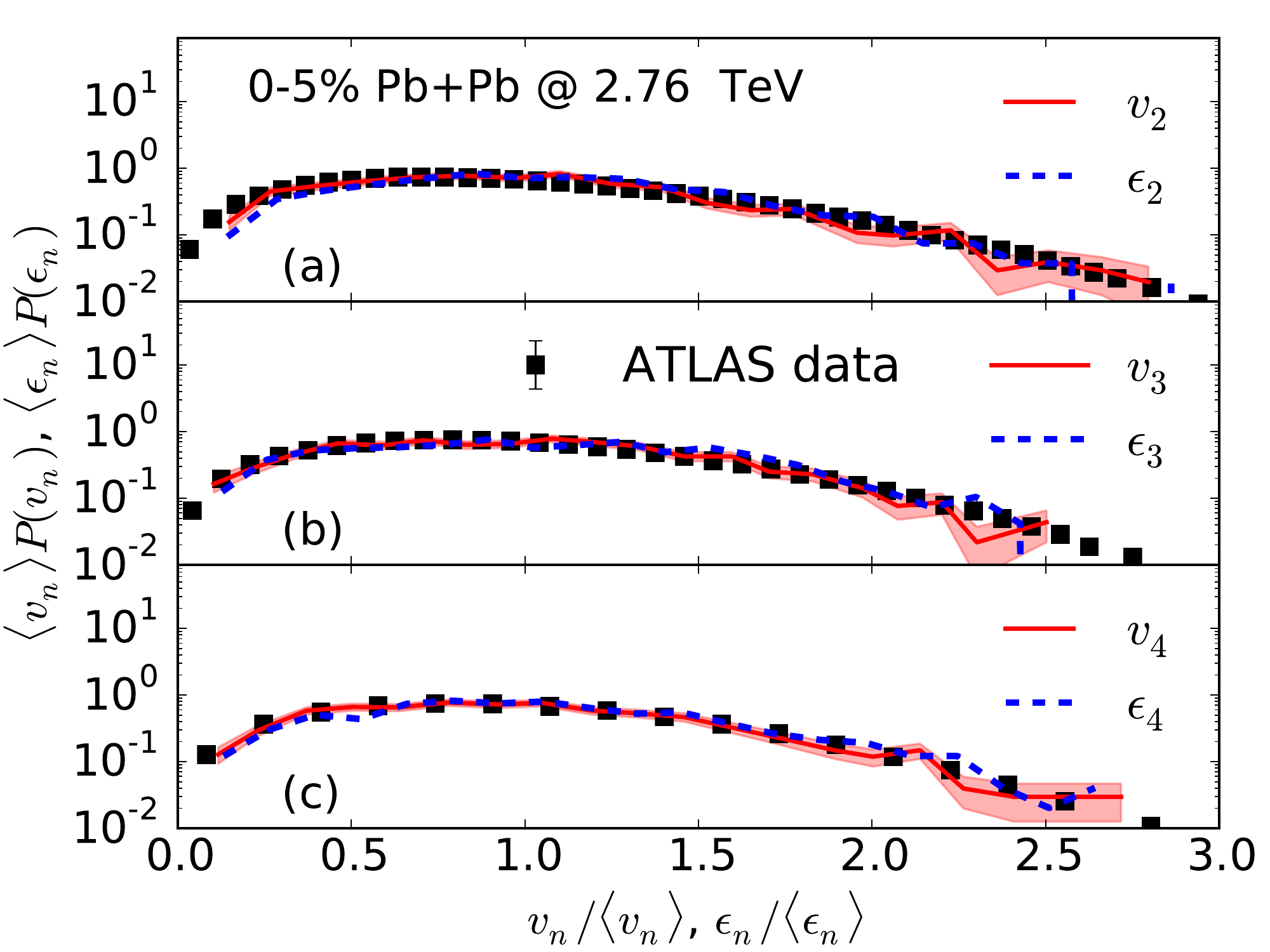} \\
  \includegraphics[width=0.9\linewidth]{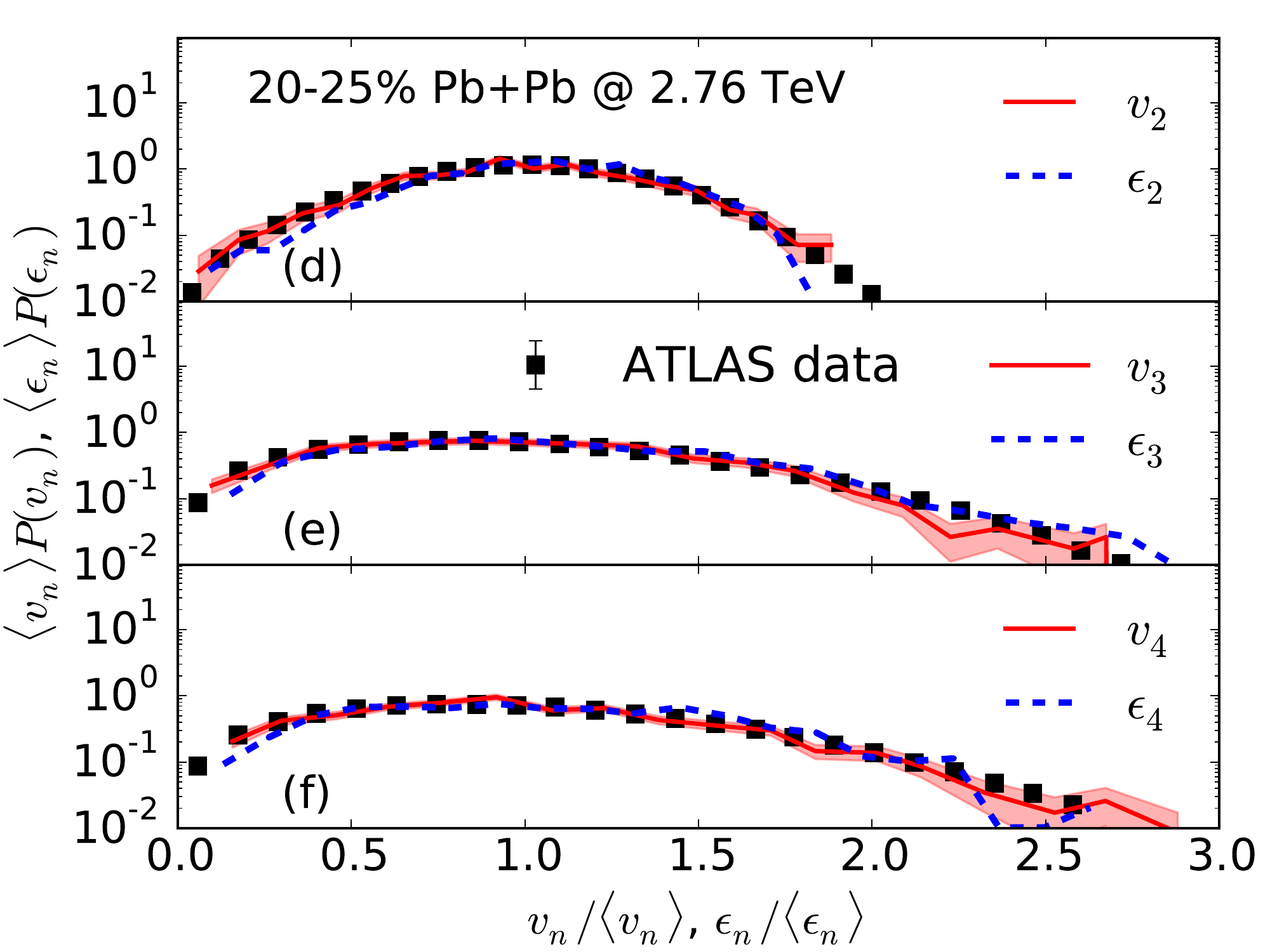} \\
   \includegraphics[width=0.9\linewidth]{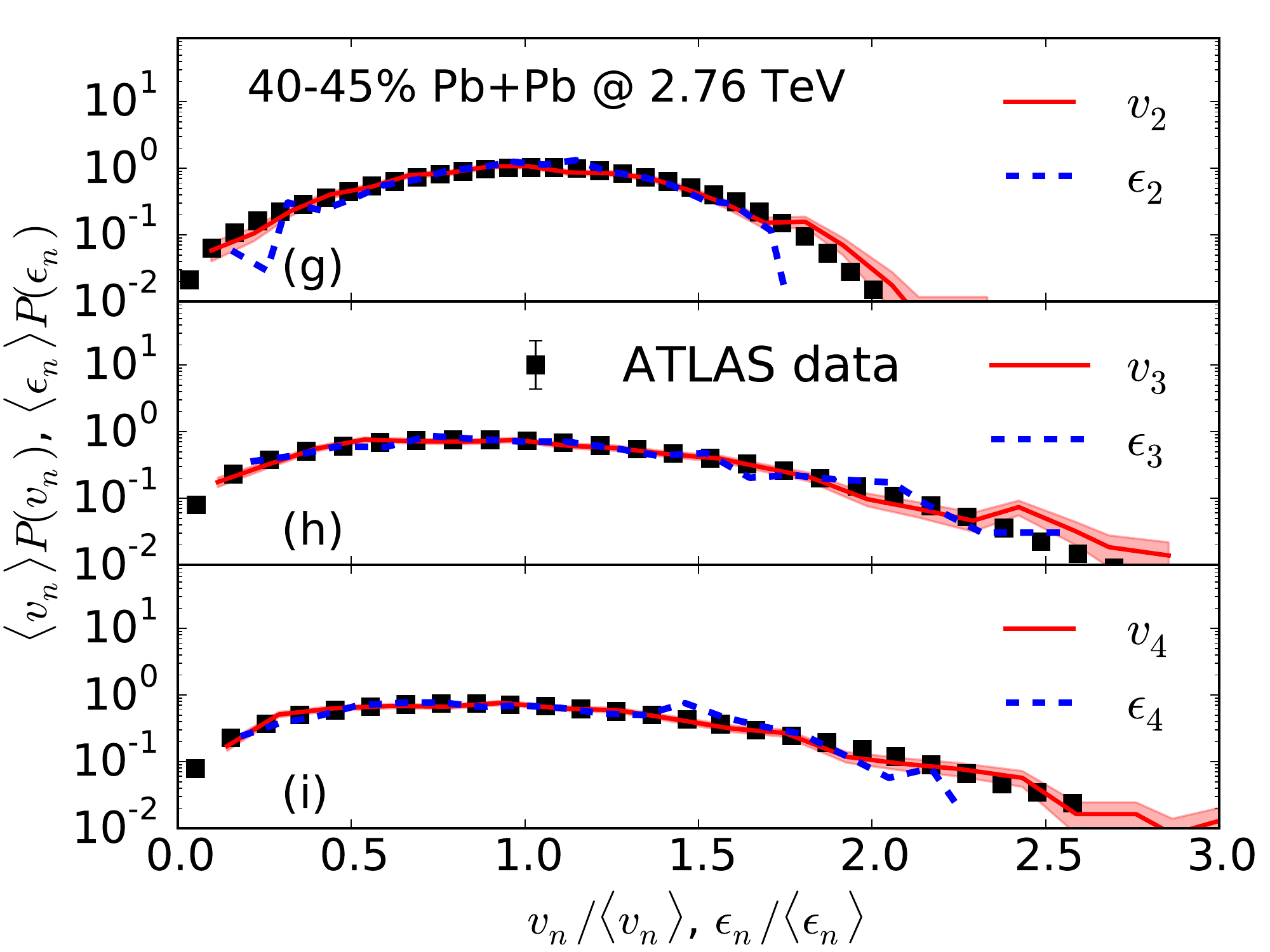} 
  \end{tabular}
  \caption{(Color online) The normalized distributions of event-by-event charged hadron anisotropic flow coefficients are compared with the measurements from the ATLAS collaboration in 0-5\% $(a$-$c)$, 20-25\% $(d$-$f)$, and 40-45\% $(g$-$i)$ centralities in Pb+Pb collisions at 2.76 TeV \cite{Aad:2013xma}. The charged hadron $v_n$ in every event are integrated from 0.5 to 3.0 GeV in transverse momentum. The normalized distributions of initial eccentricities at the end of the IP-Glasma phase ($\tau_\mathrm{0} = 0.4$ fm) are shown as dashed lines.  }
  \label{fig7}
\end{figure}
%

In Ref.~\cite{Gale:2012rq}, the authors first showed that IP-Glasma initial conditions coupled to viscous hydrodynamic simulations can reproduce the measured event-by-event distribution of charge hadron $v_n$ quite well \cite{Aad:2013xma}. In Figs.~\ref{fig7}, we show that the same quality of agreement can be achieved with a fully integrated hybrid approach which includes both shear and bulk viscosities. By comparing to the normalized distributions of initial eccentricities at the beginning of the hydrodynamic evolution, we find that the shape of the $v_n$ distributions in central to semi-peripheral centrality bins are largely determined by initial state fluctuations. Hence, in these collisions the normalized $v_n$ distributions reflect the nature of initial state fluctuations in heavy-ion collisions. In the more peripheral 40-45\% centrality, the sequential hydrodynamic evolution and the hadronic transport dynamics modifies the tail of the $v_2$ and $v_4$ distributions. This is understood as linear and cubic response in anisotropic flow harmonics \cite{Niemi:2015qia,Noronha-Hostler:2015dbi}. 
\begin{figure}[ht!]
  \centering
  \begin{tabular}{c}
  \includegraphics[width=0.9\linewidth]{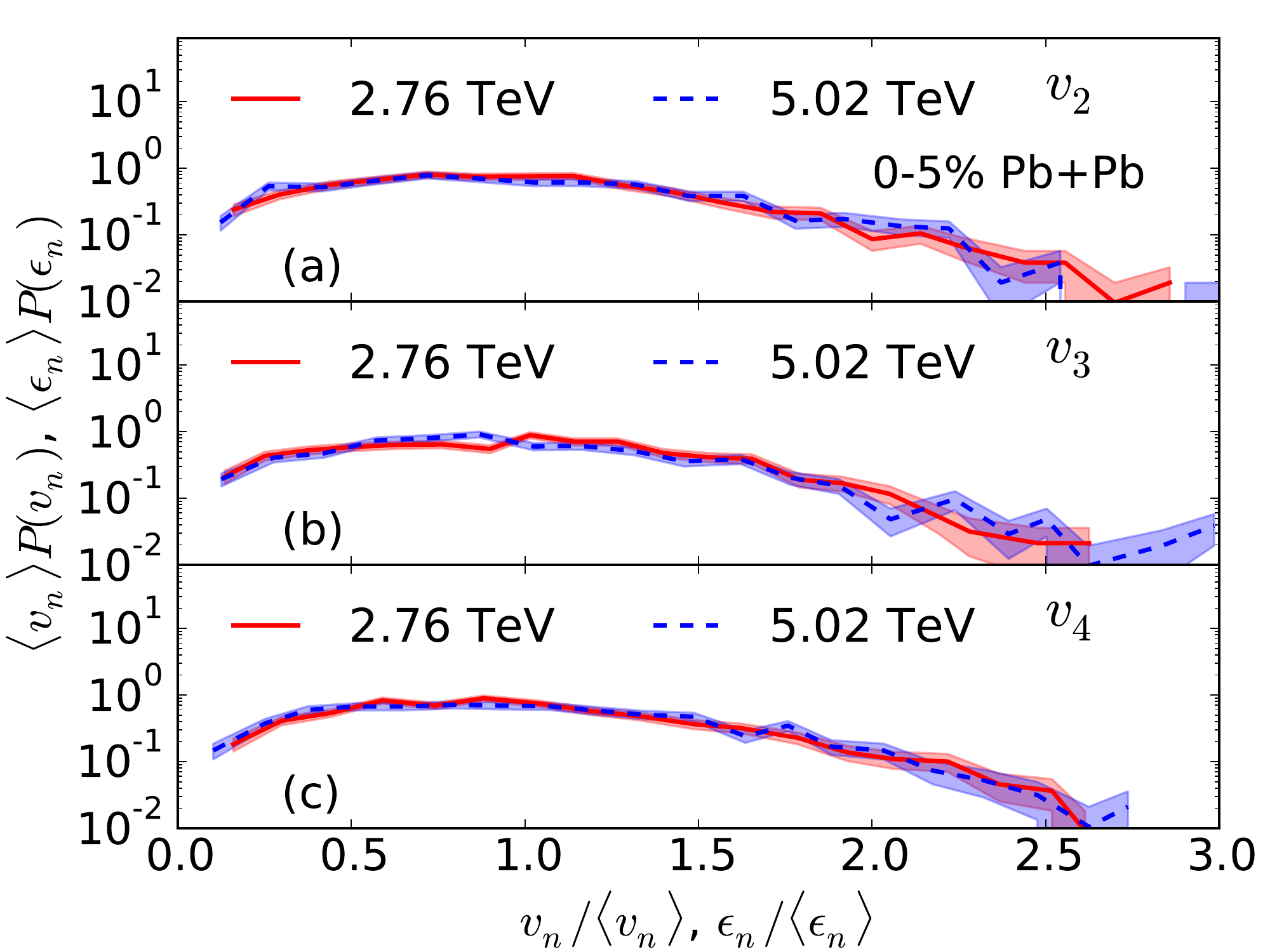} \\
  \includegraphics[width=0.9\linewidth]{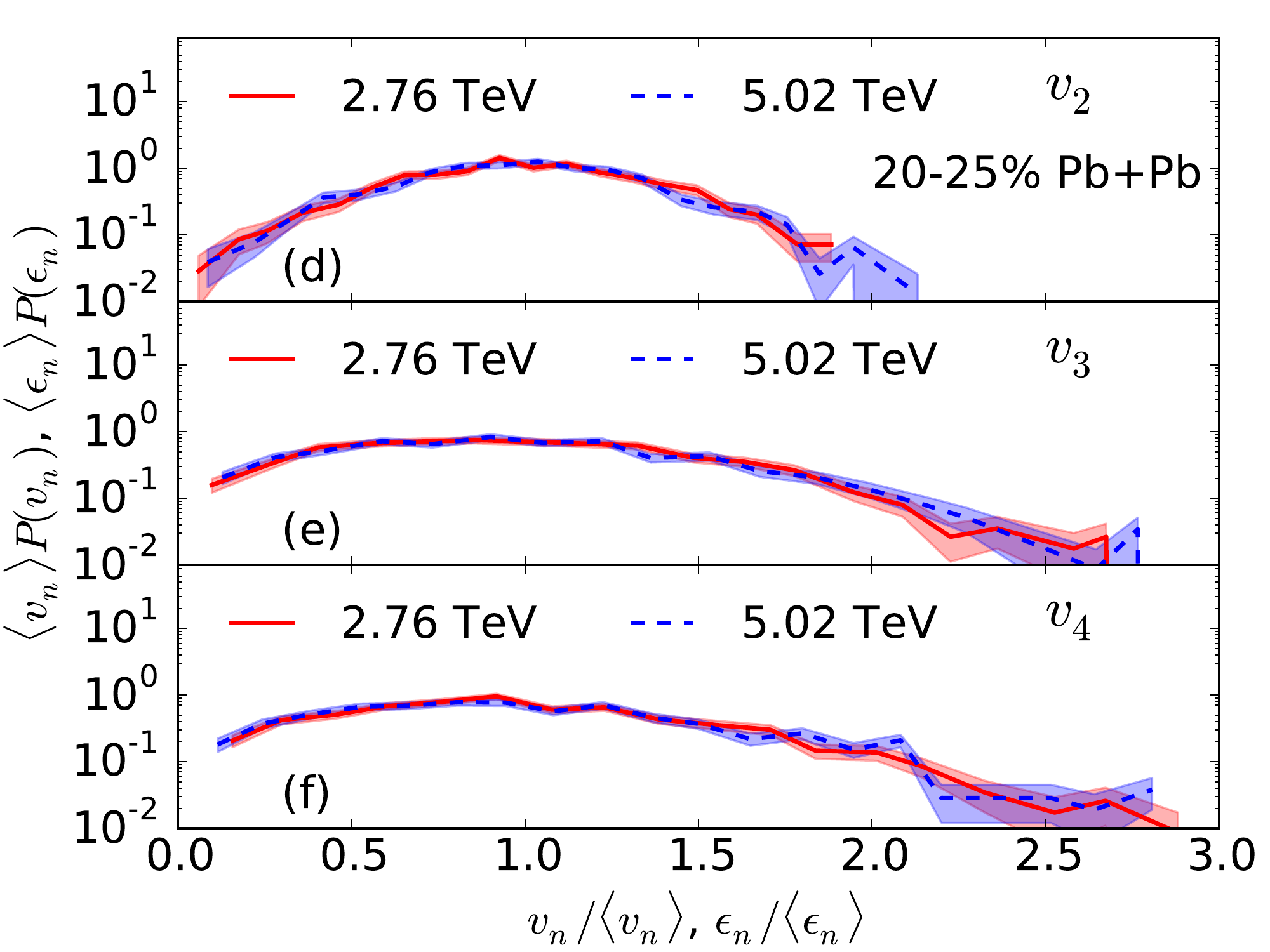} \\
  \includegraphics[width=0.9\linewidth]{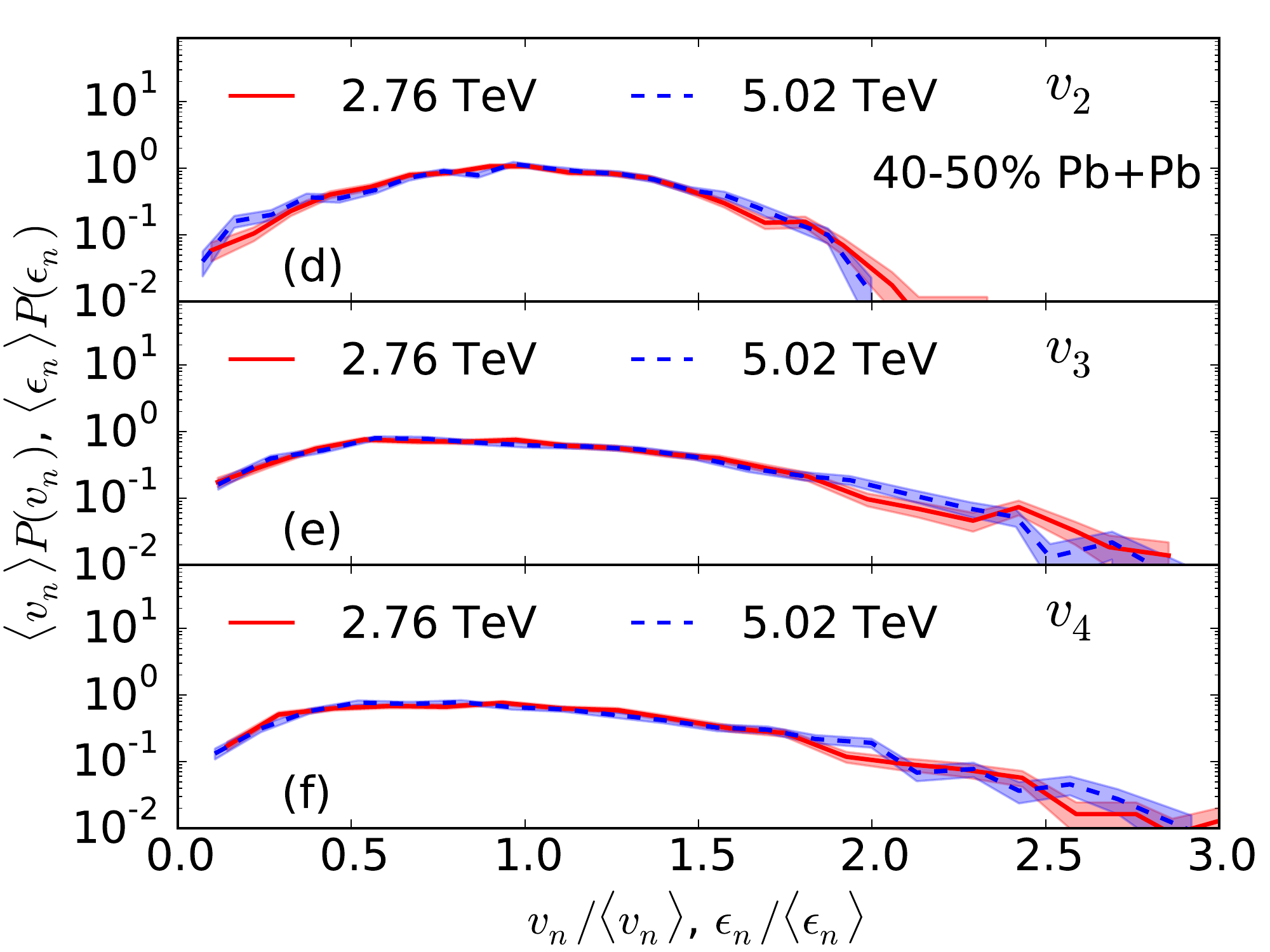}
  \end{tabular}
  \caption{(Color online) Predictions of event-by-event charged hadron $v_n$ distributions in 0-5\% and 20-25\% Pb+Pb collisions at 5.02 TeV.}
  \label{fig8b}
\end{figure}
%
Predictions of the event-by-event charged hadron $v_n$ distributions in Pb+Pb collisions at 5.02 TeV are shown in Figs.~\ref{fig8b}. The shapes of the normalized $v_n$ distributions are very close to those at 2.76 TeV. With about twice of the collision energy, the changes in the saturation scale in the IP-Glasma model are not large enough to affect the event-by-event distribution of the initial eccentricities, which imprint themselves on the final charged hadron $v_n$ distributions. This is consistent with the recent ALICE flow cumulant  measurement in Ref. \cite{Adam:2016izf}, which showed that the relative increase in $v_2\{2\}$ and $v_2\{4\}$ are the same within statistical error bars. This means that the normalized variance of the $v_2$ distribution, which can be estimated as $\sigma_{v_2}^2/\langle v_2\rangle^2 = (v_2^2\{2\} - v_2^2\{4\})/(v_2^2\{2\} + v_2^2\{4\})$, remains unchanged when the collision energy increases by 80\%.

\begin{figure*}[ht!]
  \centering
  \begin{tabular}{cc}
  \includegraphics[width=0.45\linewidth]{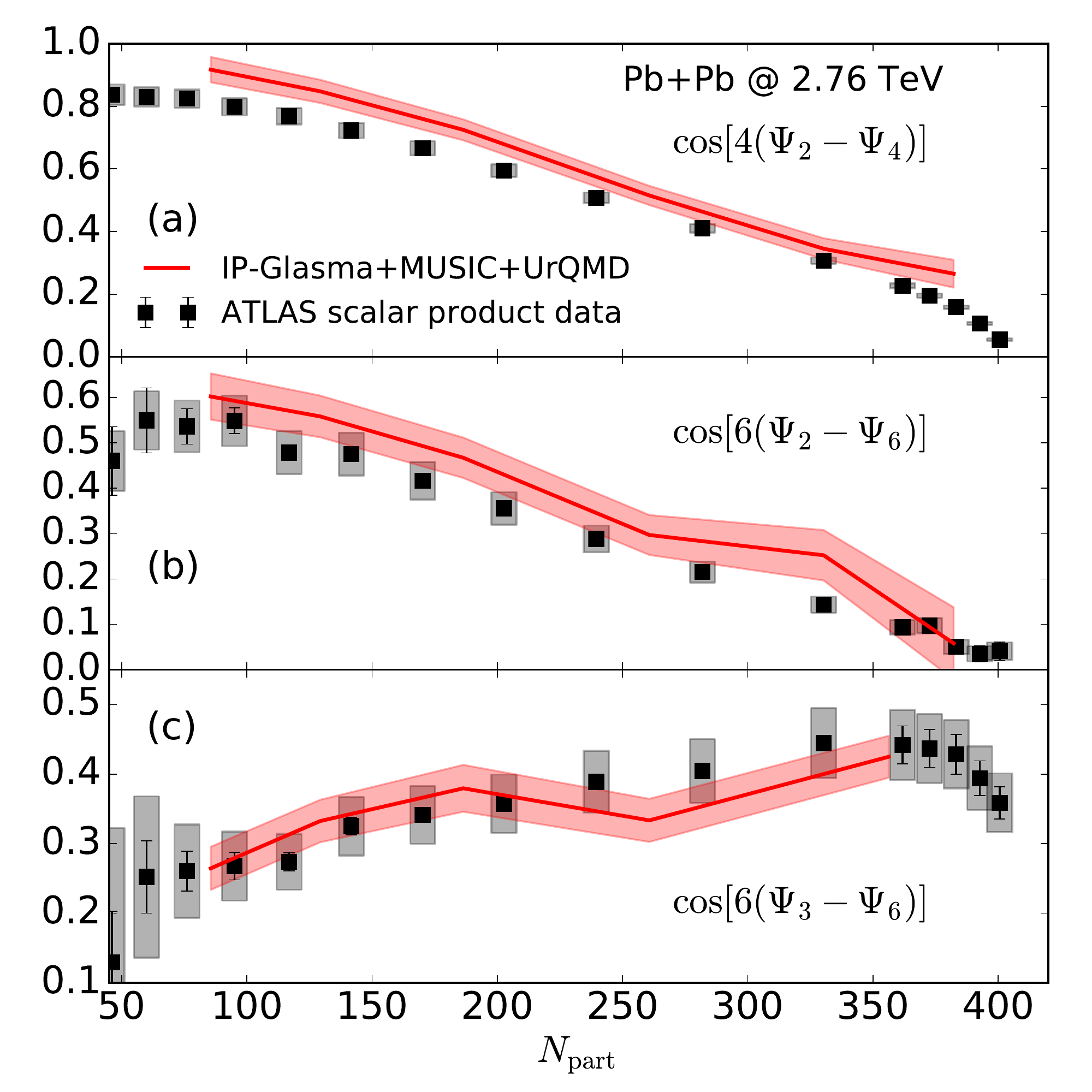} &
  \includegraphics[width=0.45\linewidth]{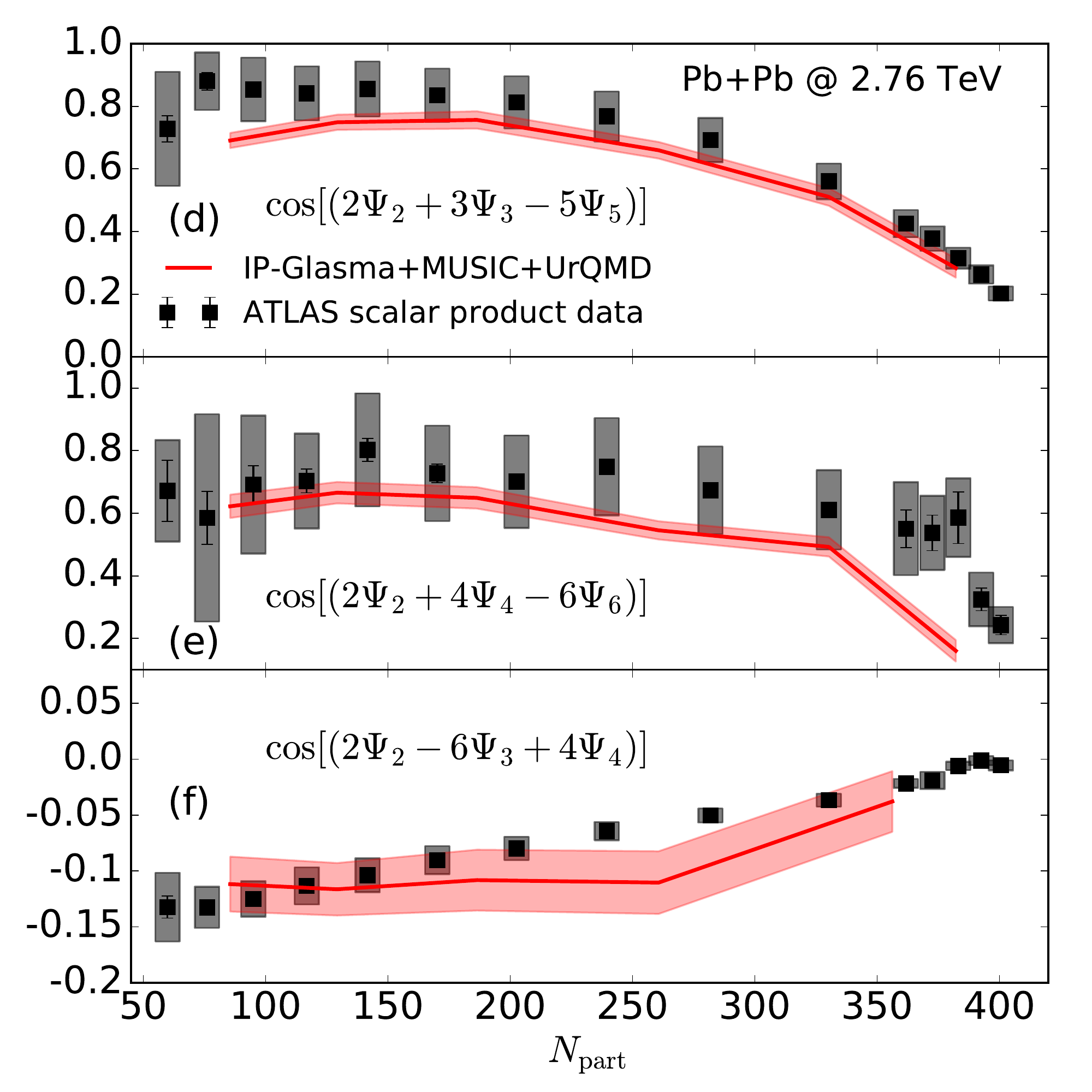}
  \end{tabular}
  \caption{(Color online) The charged hadron event-plane correlations are compared with the ATLAS measurements \cite{Aad:2014fla} using the scalar-product method in Pb+Pb collisions at 2.76 TeV. The values of $N_{part}$ at different centrality bins are estimated using MC-Glauber model according to Table 1 in Ref. \cite{Aad:2014fla}. }
  \label{fig9}
\end{figure*}
%

Event-by-event fluctuations in heavy-ion collisions also result in non-trivial correlations among different orders of harmonic flow coefficients. In Figs.~\ref{fig9}, we compute the two and three event-plane correlations from our hybrid simulations and compare them with the ATLAS measurements in Pb+Pb collisions at 2.76 TeV \cite{Aad:2014fla}. We used the scalar-product method to compute these correlators \cite{Jia:2014jca},
\begin{eqnarray}
&& \cos\left[c_1 n_1 \Psi_{n_1} - c_2 n_2 \Psi_{n_2} \right] \notag \\
&& \qquad\qquad = \frac{{\rm Re}\{\langle {\bf Q}_{n_1}^{c_1} ({\bf Q}_{n_2}^{c_2})^* \rangle \}}{\sqrt{\langle {\bf Q}_{n_1}^{c_1} ({\bf Q}_{n_1}^{c_1})^* \rangle} \sqrt{\langle ({\bf Q}_{n_2}^{c_2}) ({\bf Q}_{n_2}^{c_2})^* \rangle}}
\end{eqnarray}
and
\begin{eqnarray}
&&\cos\left[c_1 n_1 \Psi_{n_1} + c_2 n_2 \Psi_{n_2} - c_3 n_3 \Psi_{n_3} \right] \notag \\
&& = \frac{{\rm Re} \{ \langle {\bf Q}_{n_1}^{c_1} {\bf Q}_{n_2}^{c_2}({\bf Q}_{n_3}^{c_3})^* \rangle \}}{\sqrt{\langle {\bf Q}_{n_1}^{c_1} ({\bf Q}_{n_1}^{c_1})^* \rangle \langle ({\bf Q}_{n_2}^{c_2}) ({\bf Q}_{n_2}^{c_2})^* \rangle \langle ({\bf Q}_{n_3}^{c_3}) ({\bf Q}_{n_3}^{c_3})^* \rangle}},
\end{eqnarray}
where the particle flow vector ${\bf Q}_n$ is defined in Eq. (\ref{eq.Qn}). The imaginary part of the correlation function goes to zero after the event average \cite{Heinz:2013bua}. 
In the 6 panels of Fig.~\ref{fig9}, our results show fairly good agreement with the experimental measurement as a function of centrality. In the panels $(c)$ and $(f)$, the event-plane correlators $\cos[6(\Psi_3 - \Psi_6)]$ and $\cos[2\Psi_2 - 6\Psi_3 + 4\Psi_4]$ require larger statistics compared to the others. We combined the events in central 0-5\% and 5-10\% bins to 0-10\% to reduce the statistical noise in our results. 
\begin{figure*}[ht!]
  \centering
  \begin{tabular}{cc}
  \includegraphics[width=0.45\linewidth]{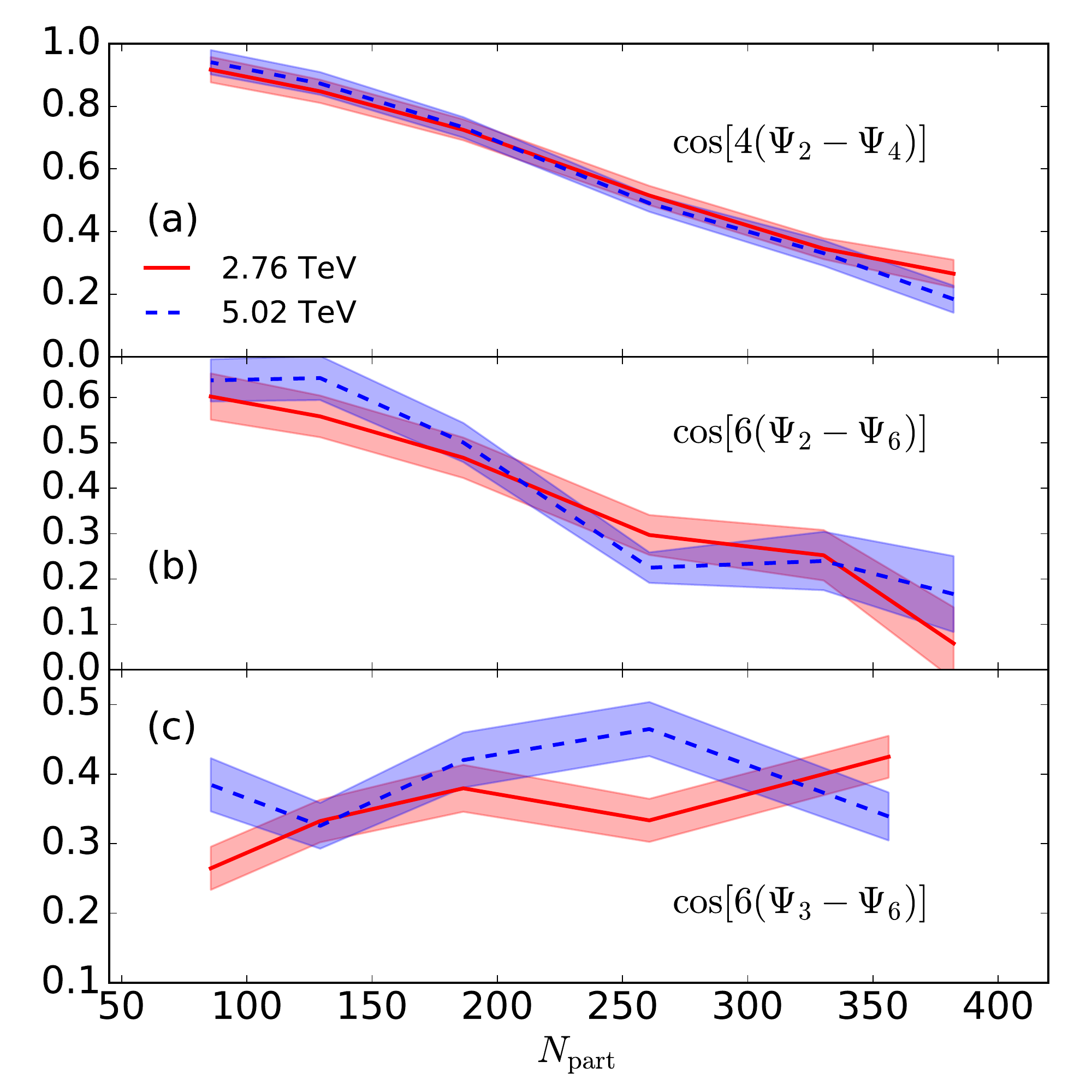} &
  \includegraphics[width=0.45\linewidth]{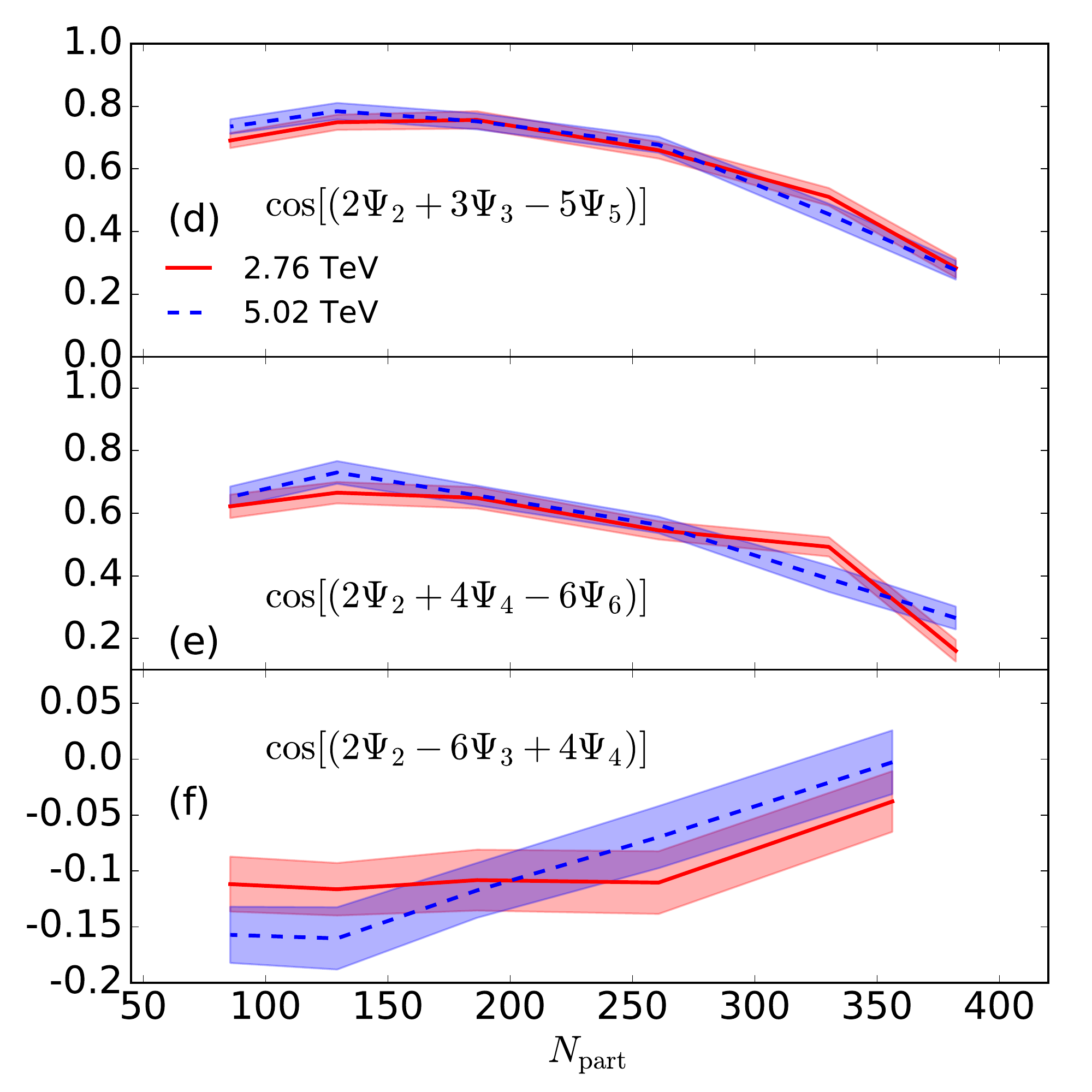}
  \end{tabular}
  \caption{(Color online) Predictions of charged hadron event-plane correlations in Pb+Pb collisions at 5.02 TeV.}
  \label{fig10}
\end{figure*}
%
In Fig.~\ref{fig10}, predictions of the flow event-plane correlations at 5.02 TeV are shown. Compared with the results at 2.76 TeV, the correlation strengths are similar at higher collision energies. This reflects that the hydrodynamic  response is very similar in the two collision energies. 

\begin{figure*}[ht!]
  \centering
  \includegraphics[width=0.9\linewidth]{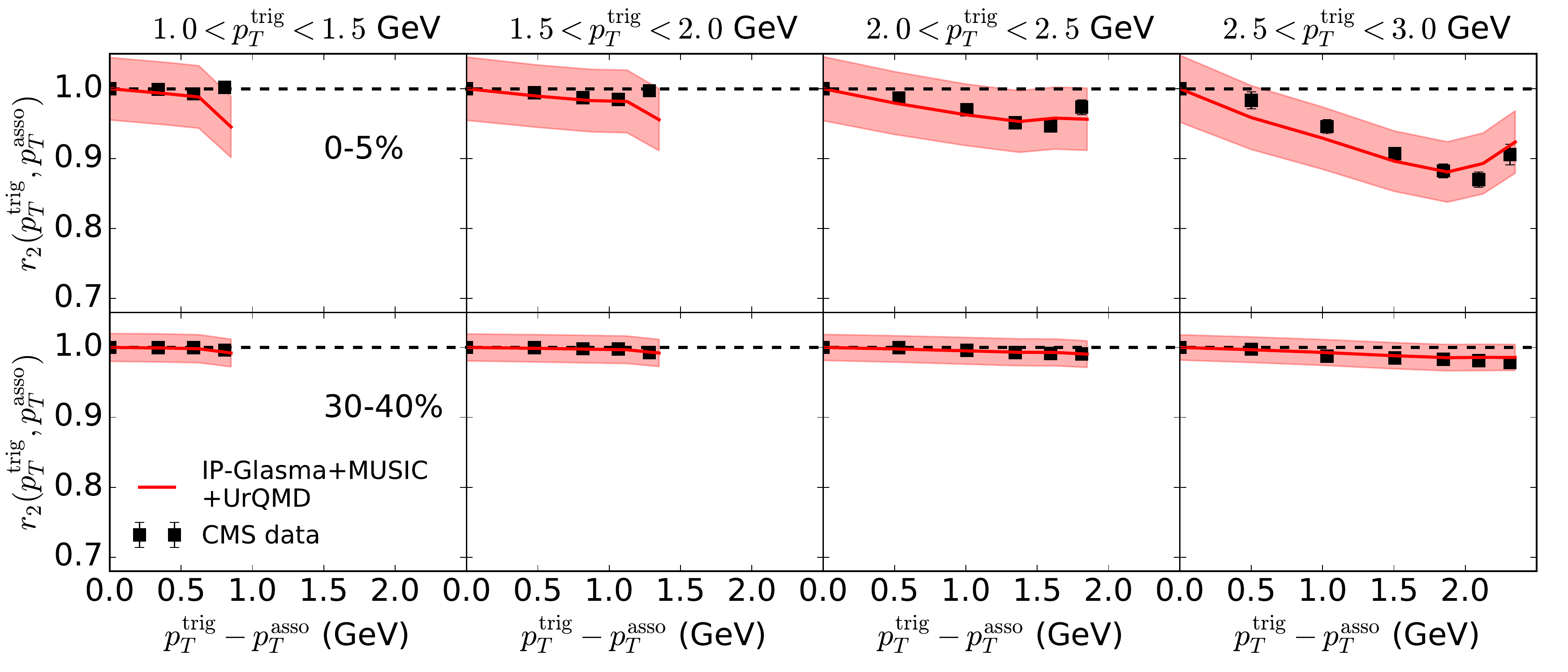}
  \includegraphics[width=0.9\linewidth]{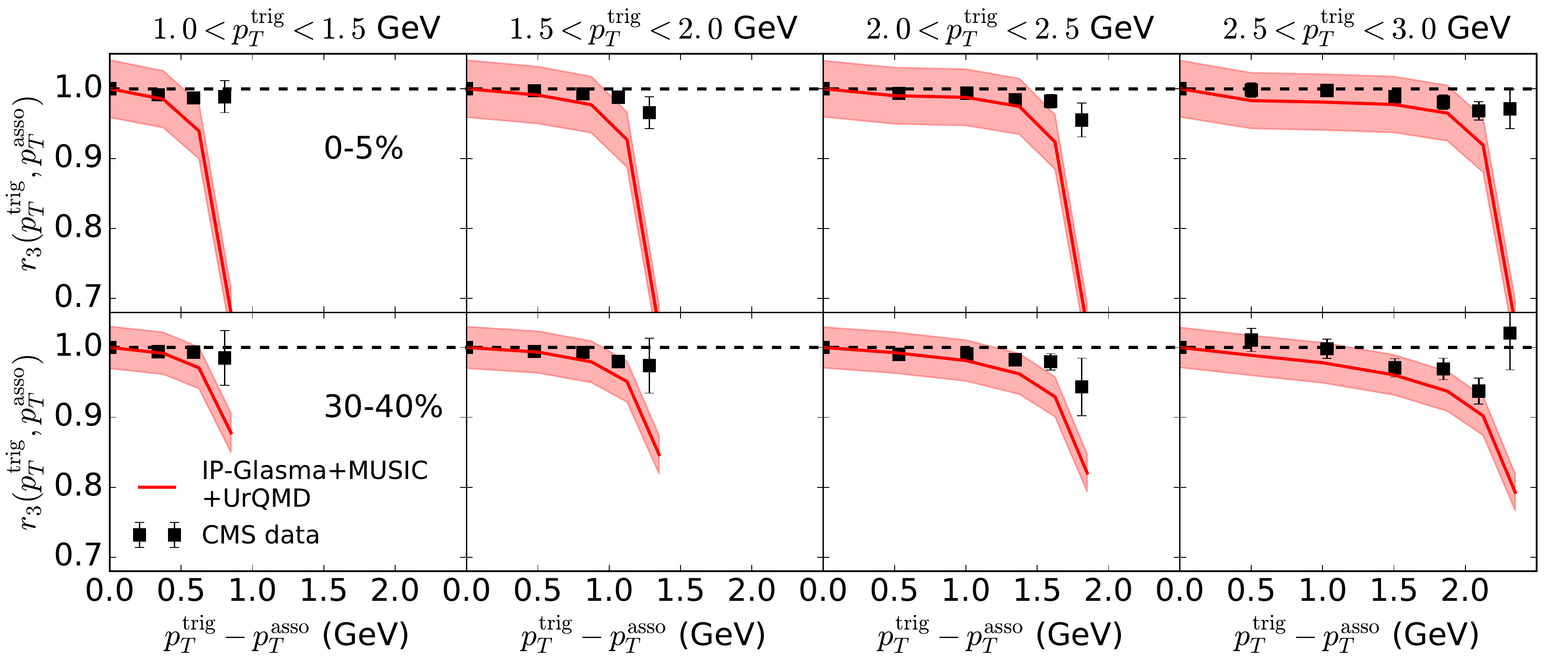}
  \caption{(Color online) The charged hadron flow factorization breaking ratio $r_n$ are compared with the CMS measurements \cite{Khachatryan:2015oea} in Pb+Pb collisions at 2.76 TeV. }
  \label{fig11}
\end{figure*}
%

Figs.~\ref{fig11} further show the flow factorization breaking ratios $r_{2,3}$ in Pb+Pb collision at 2.76 TeV compared to the CMS measurements \cite{Khachatryan:2015oea}. This $r_n$ ratios reflect the correlations of $v_n(p_T)$ in different $p_T$ bins,
\begin{eqnarray}
&&r_n(p^\mathrm{trig}_T, p^\mathrm{asso}_T) \notag \\
&&= \frac{{\rm Re} \{\langle {\bf Q}_n(p^\mathrm{trig}_T) ({\bf Q}_n(p^\mathrm{asso}_T))^* \rangle\} }{\sqrt{\langle {\bf Q}_n(p^\mathrm{trig}_T) {\bf Q}^*_n(p^\mathrm{trig}_T) \rangle \langle {\bf Q}_n(p^\mathrm{asso}_T) {\bf Q}^*_n(p^\mathrm{asso}_T) \rangle}}.
\end{eqnarray}
Our hybrid calculations reproduce fairly well the CMS measured $r_{2,3}(p^\mathrm{trig}_T, p^\mathrm{asso}_T)$ in both central and semi-peripheral centrality bins. A sharp drop in the $r_3$ ratio is found in the theoretical calculation at large values of $p^\mathrm{trig}_T - p^\mathrm{asso}_T$. About 75\% of the factorization breaking in this $p_T$ bin can be traced back to the flow angle decorrelation between $\Psi_3(p^\mathrm{trig}_T)$ and $\Psi_3(p^\mathrm{asso}_T)$. 
\begin{figure*}[ht!]
  \centering
  \includegraphics[width=0.9\linewidth]{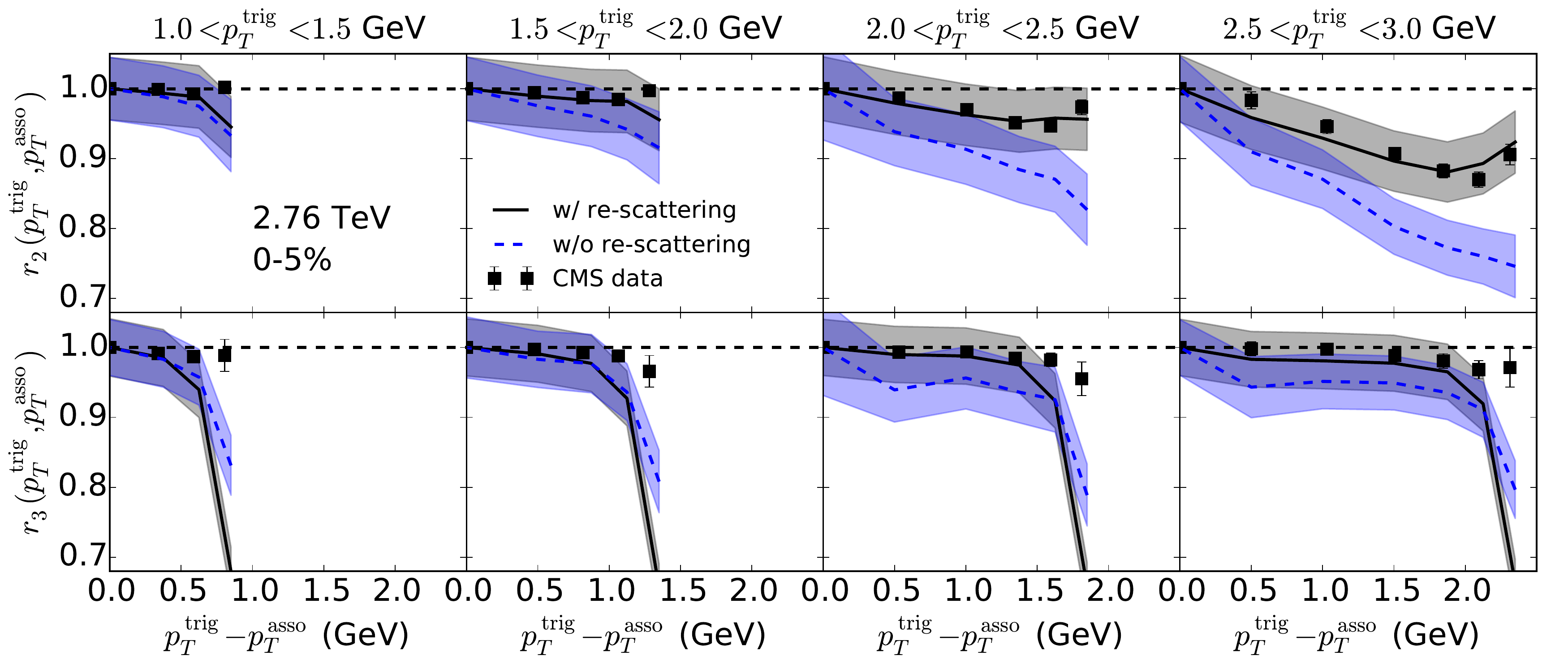}
  \caption{(Color online) The effect of hadronic rescatterings on the flow factorization breaking ratios $r_2$ and $r_3$.
}
  \label{fig.rn.urqmd}
\end{figure*}
\begin{figure*}[ht!]
  \centering
  \includegraphics[width=0.9\linewidth]{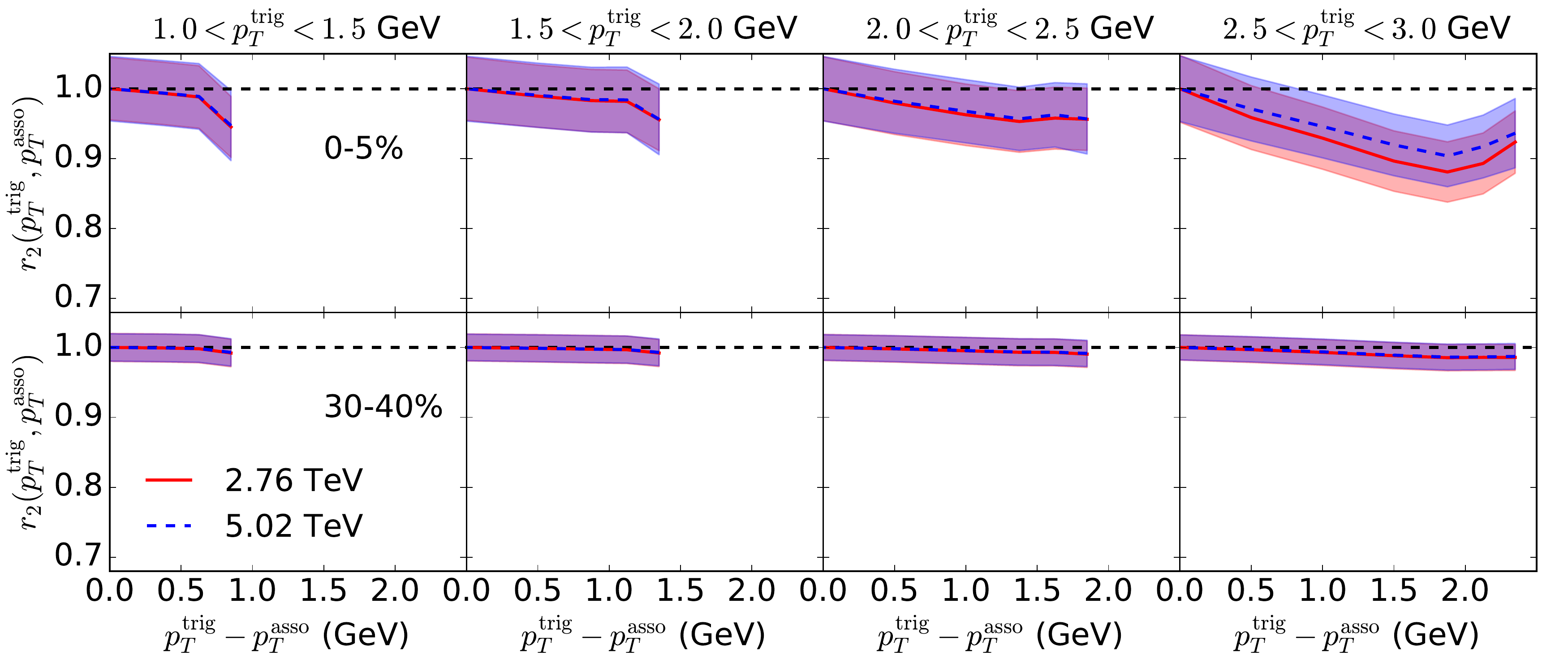}
  \includegraphics[width=0.9\linewidth]{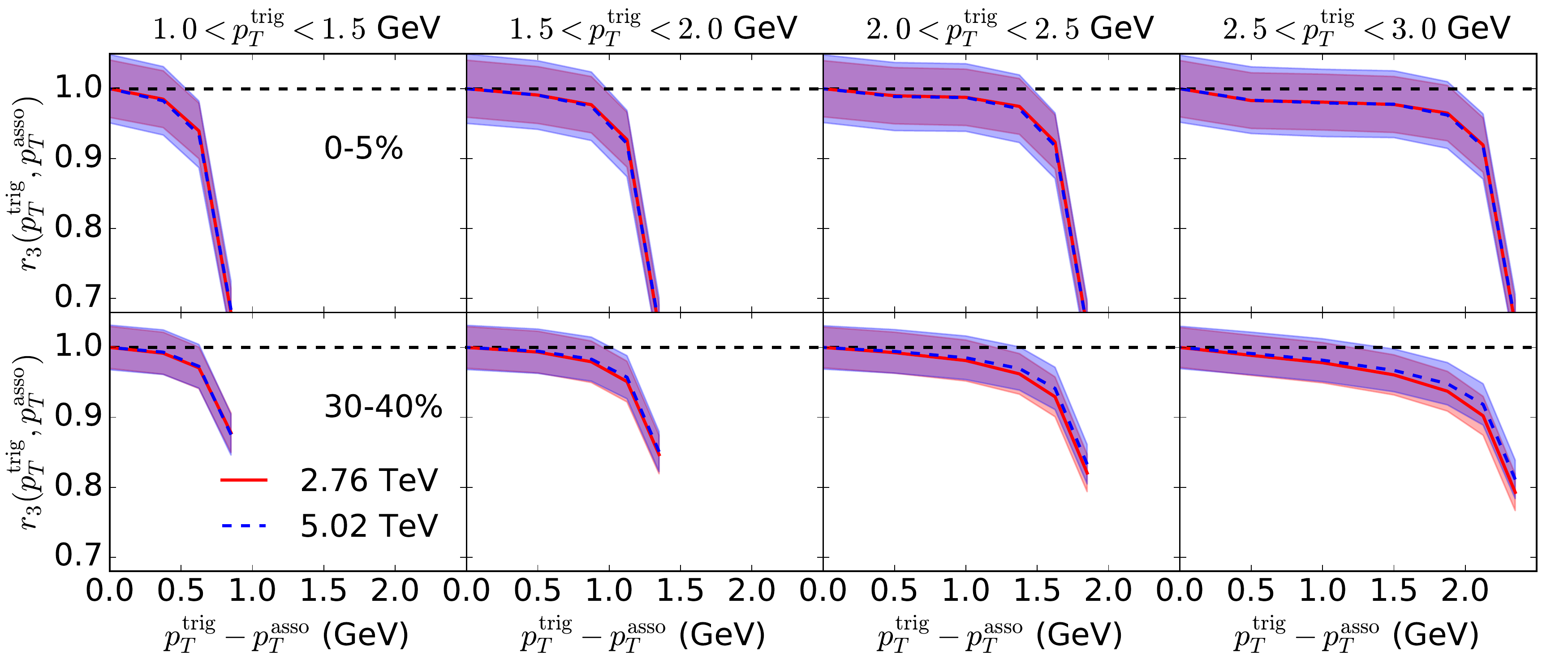}
  \caption{(Color online) Predictions of charged hadron flow factorization breaking ratio in Pb+Pb collisions at 5.02 TeV. 
}
  \label{fig12}
\end{figure*}
%
Fig.~\ref{fig.rn.urqmd} illustrates the effects of hadronic rescatterings on the flow factorization breaking ratios. For the $r_2$ ratio, the scattering among hadrons increases the correlations of $v_2(p_T)$ at different $p_T$ bins. Situation becomes a little bit complicated for the $r_3$ ratio. The additional hadronic scatterings strengthen the correlations for $v_3(p_T)$ among different $p_T$ bins with $p_T > 0.5$ GeV. For $p_T < 0.5$ GeV (the largest value of $p_T^\mathrm{trig} - p_T^\mathrm{asso}$), hadronic scatterings randomize the $v_3$ flow angle. This randomization seems not present for $n= 2$ but is more severe for high order of $n$. 
Finally, Figs.~\ref{fig12} predicts the $r_{2,3}$ ratio at the new 5.02 TeV. The magnitude of factorization breaking is about the same at both collision energies.

\section{Conclusions}

In this work, we provided quantitative predictions for hadronic flow observables in Pb+Pb collisions at the new 5.02 TeV collision energy of the LHC. As a starting point of our predictions, we demonstrated that our hybrid framework provided excellent descriptions of a variety of current existing flow measurements in Pb+Pb collisions at 2.76 TeV. The anisotropic flow results of extrapolating to 5.02 TeV showed excellent agreement with the recent ALICE measurements. Such a successful postdiction, together with the fact that the same effective shear viscosity was used at two collision energies, suggest the temperature dependence of $\eta/s(T)$ is rather mild above $T_c$. Detailed predictions of identified particle observables are provided. Future comparison with experimental data can help us to better understand the dynamical evolution of the collision systems, especially the dynamics in the hadronic phase. Event-by-event distributions of anisotropic flow coefficients and flow correlations among them are studied at the two collision energies of the LHC. The $p_T$ dependence and centrality dependence of these correlation observables remain qualitatively the same at 5.02 TeV compared to those measured at 2.76 TeV. The small quantitative changes can help to test the variation of hydrodynamic response at higher collision energies. 

In future work, we will embed QCD jet showers and mini-jets contribution into our hydrodynamic medium. The results of this approach can improve the description of particle spectra and flow anisotropy coefficients at intermediate $p_T$ regions, $p_T \gtrsim 2$ GeV. The violation of the mass ordering of multi-strange hadrons in the hybrid calculations needs a more detailed study in the hadronic transport model. With the calibrated medium, studies on penetrating observables, such as direct photons and energy loss of QCD jets, will be presented in a forthcoming publication.

\appendix*
\section{The effects of out-of-equilibrium correction $\delta f$ on flow observables}

In this work, the form of out-of-equilibrium correction $\delta f$ in the Cooper-Frye freeze-out is chosen as this derived from Boltzmann equation assuming relaxation time approximation. The exact form of $\delta f$ for shear and bulk viscosities are still unknown. Moreover, because the $\delta f$ correction increases with $p_T$, higher order corrections are more important and can not be neglected at high $p_T$. So $\delta f$ corrections introduce theoretical uncertainties in our results. In this appendix, we would like to study the sensitivity of the $\delta f$ corrections on hadronic flow observables. 

\begin{figure}[ht!]
  \centering
  \includegraphics[width=1\linewidth]{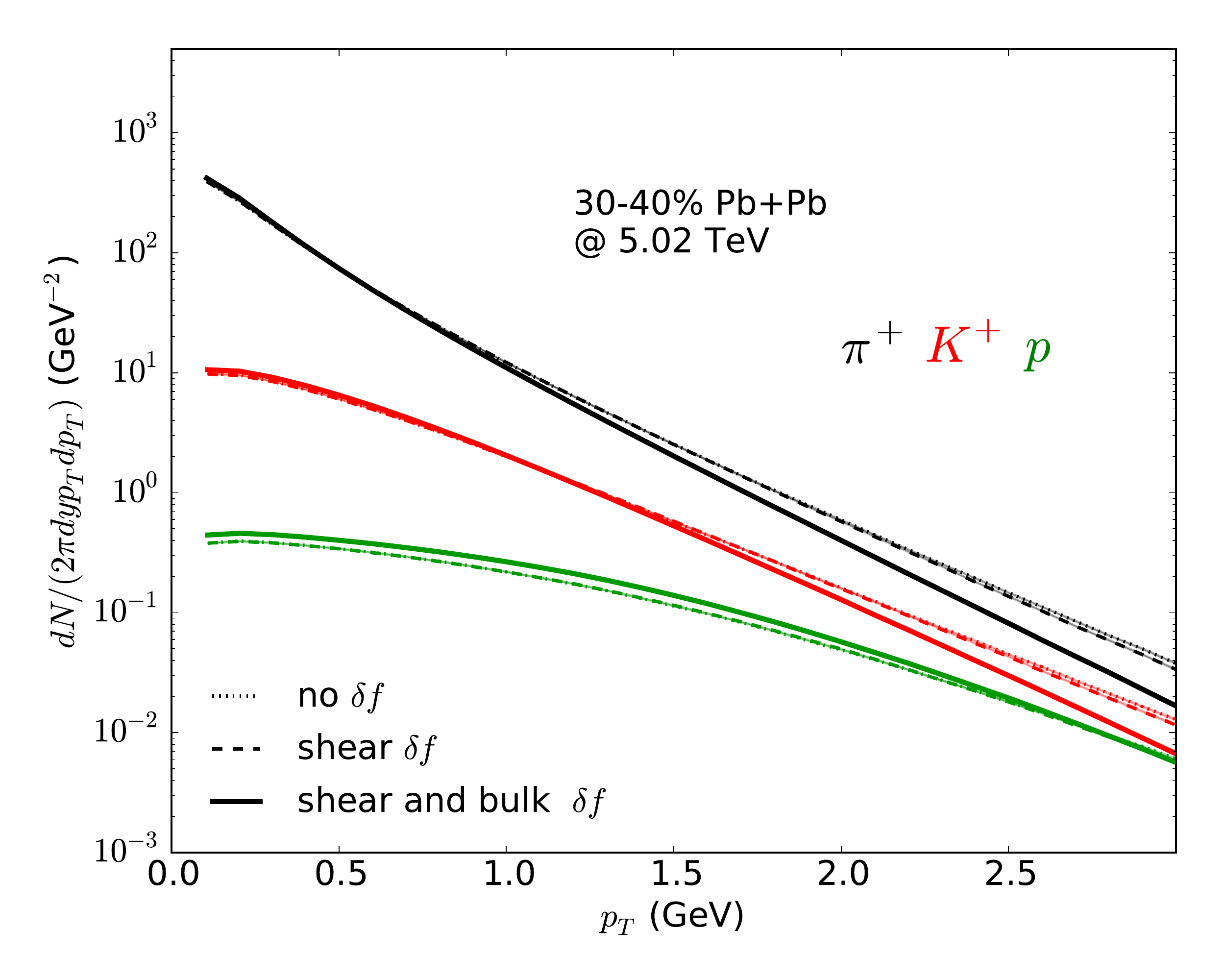}
  \caption{(Color online) Identified particle spectra with and without $\delta f$ corrections from shear and bulk viscosities.}
  \label{fig18}
\end{figure}
%
\begin{figure*}[ht!]
  \centering
  \begin{tabular}{cc}
  \includegraphics[width=0.4\linewidth]{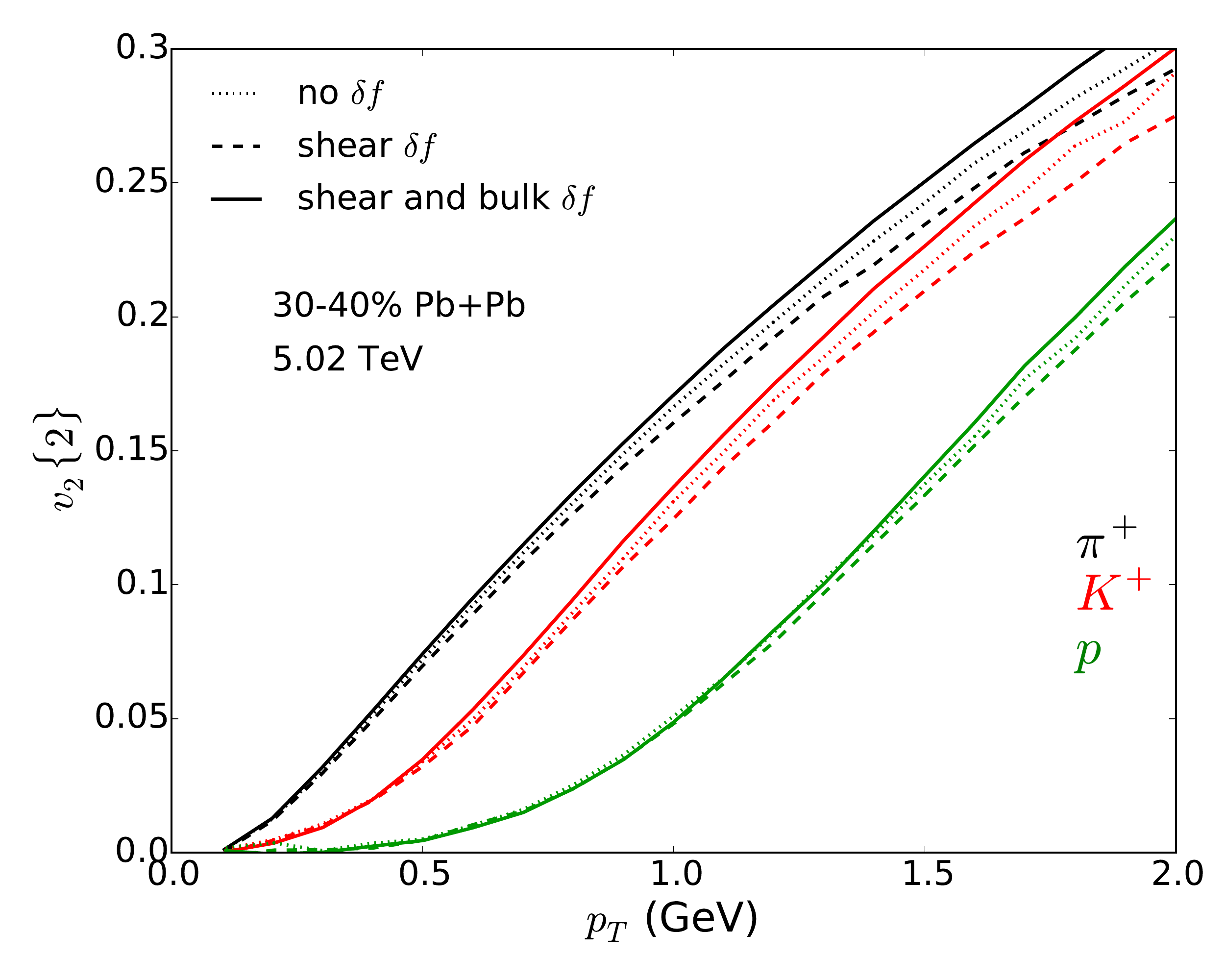} &
  \includegraphics[width=0.4\linewidth]{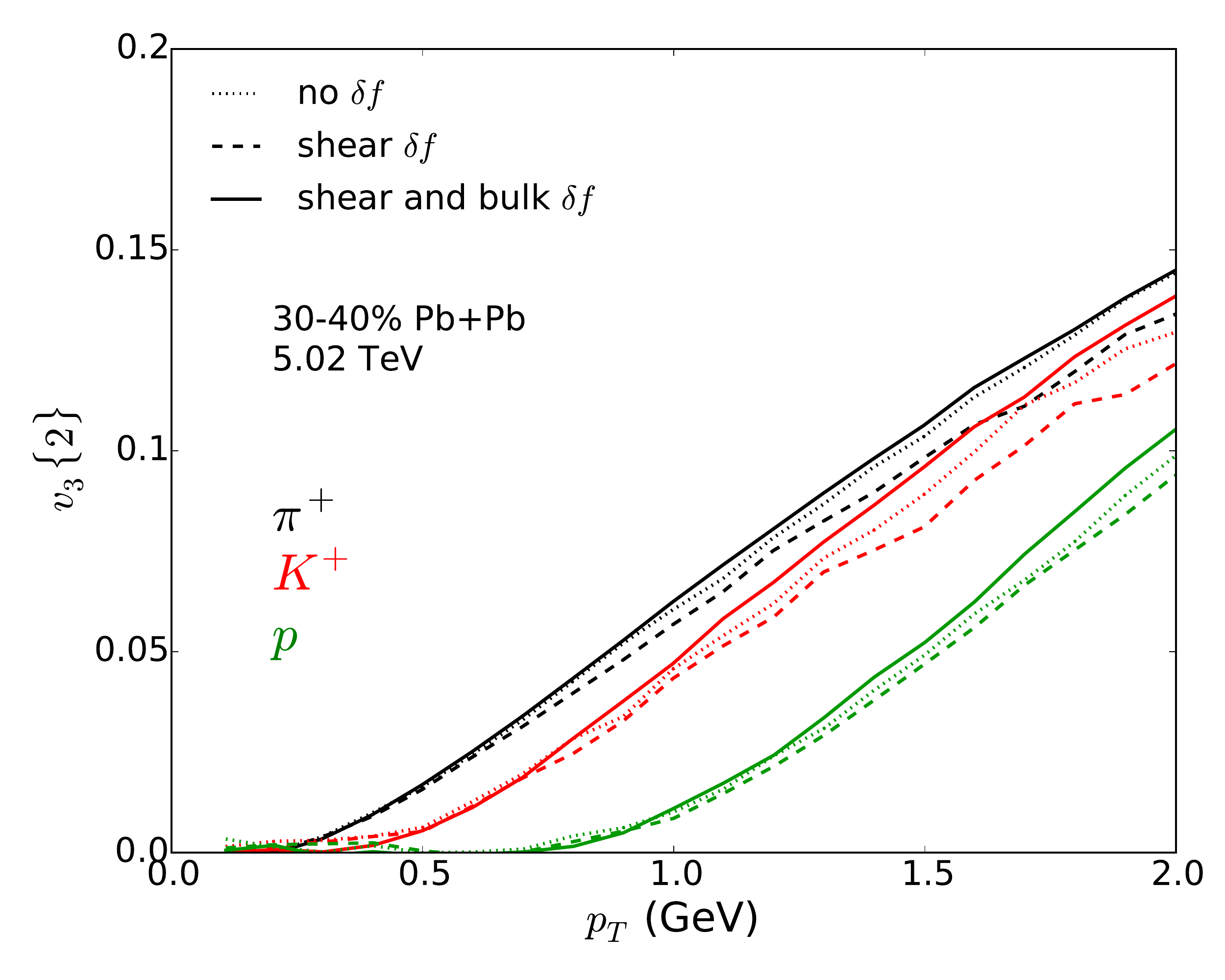}
  \end{tabular}
  \caption{(Color online) Identified particle $p_T$-differential $v_2(p_T)$ and $v_3(p_T)$ with and without the shear and bulk $\delta f$ corrections. }
  \label{fig19}
\end{figure*}
%

Fig.~\ref{fig18} shows the effect of shear and bulk $\delta f$ on identified particle $p_T$ spectra. The shear $\delta f$ only modifies the particle spectra slightly. This is because the specific shear viscosity $\eta/s = 0.095$ is relatively small in our calculation. The size of shear stress tensor $\pi^{\mu\nu}$ is small at the freeze-out. Meanwhile, the out-of-equilibrium correction from bulk viscosity plays an important role. The bulk $\delta f$ steepens the particle spectra. The expression for bulk $\delta f$, given in Eq. (\ref{eq.bulkdeltaf}), has two terms: a mass dependent term that increases particle yield at low energy and a negative term that increases linearly with energy. The transition from positive to negative correction happens at momentum $p = m\sqrt{3c_s^2/(1 - 3c_s^2)}$. The pion and kaon spectra are suppressed for $p_T > 1.5$ GeV. For heavier protons, this transition happens at $\sim$2.5 GeV. The low $p_T$ proton spectra are enhanced which leads to an increase in the particle yield.

The $\delta f$ effects on particle $p_T$-differential $v_{2,3}(p_T)$ are studied in Fig.~\ref{fig19}. The shear $\delta f$ suppresses $v_n(p_T)$ while the bulk $\delta f$ increases it. Because the bulk $\delta f$ correction is isotropic, the increase of $v_n(p_T)$ is because the particle spectra are strongly suppressed by the $\delta f^\mathrm{bulk}$. The magnitude of bulk $\delta f$ correction is larger than the shear correction. So the identified particle $v_n(p_T)$ increases with $\delta f$ corrections.  

\begin{figure*}[ht!]
  \centering
  \includegraphics[width=0.9\linewidth]{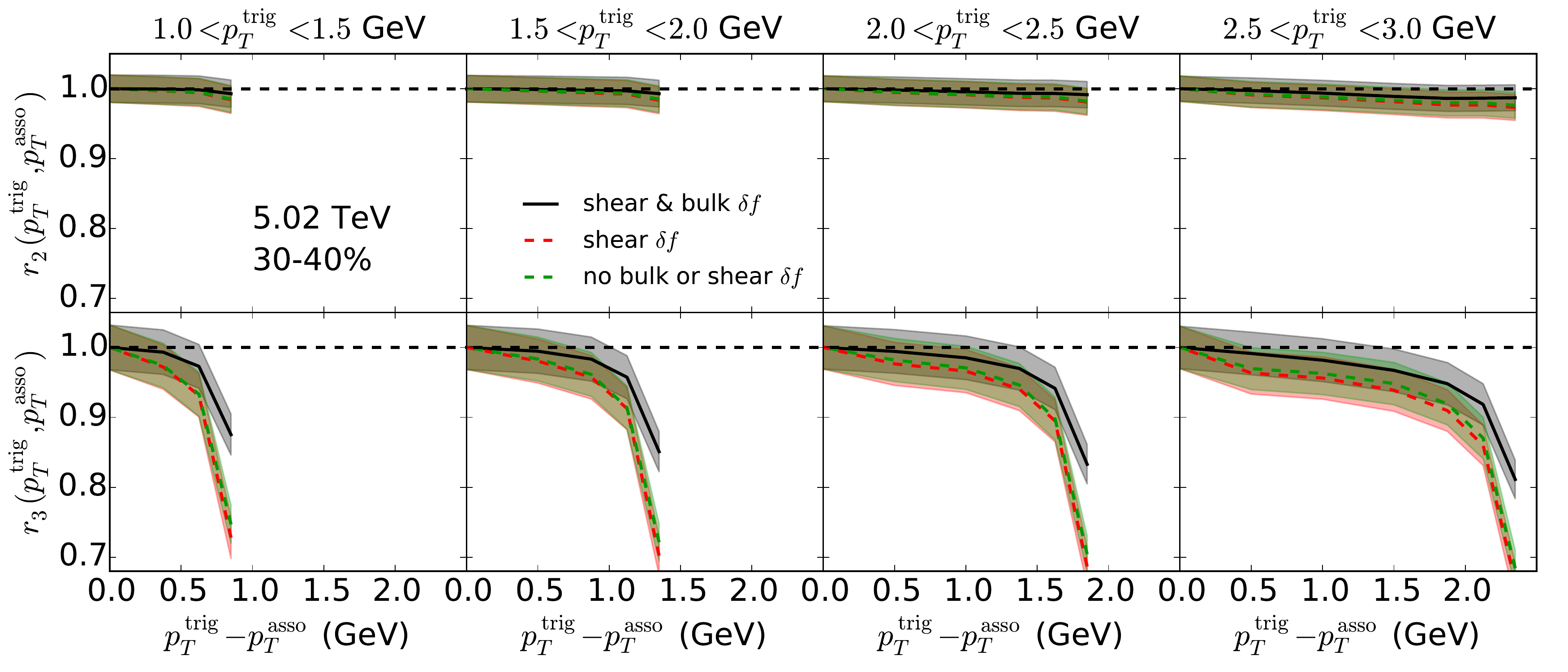}
  \caption{(Color online) The effect of shear and bulk $\delta f$ on the flow factorization breaking ratios $r_2$ and $r_3$.
}
  \label{fig20}
\end{figure*}
In Fig.~\ref{fig20}, we study the effects of shear and bulk $\delta f$ on the flow factorization breaking ratios $r_2$ and $r_3$. Both shear and bulk $\delta f$ has negligible effects on the $r_2$ ratio in all $p_T$ bins. The bulk viscous $\delta f$ suppresses the $r_3$ ratio at large difference between $p_T^\mathrm{trig}$ and $p_T^\mathrm{asso}$. This means that the triangular flow correlation between small and large $p_T$ bins is reduced by the bulk viscous $\delta f$.

\begin{acknowledgements}
This work was supported in part by the Natural Sciences and Engineering Research Council of Canada. 
Computations were made in part on the supercomputer Guillimin from McGill University, managed by Calcul Qu\'ebec and Compute Canada. The operation of this supercomputer is funded by the Canada Foundation for Innovation (CFI), NanoQu\'ebec, RMGA and the Fonds de recherche du Qu\'ebec - Nature et technologies (FRQ-NT). Finally, we would like to thank Bj\"oern Schenke for insightful discussions and for providing his IP-Glasma code for detailed numerical cross checking.

\end{acknowledgements}

\bibliography{references}

\end{document}